\documentclass[12pt]{iopart}
\usepackage{iopams}

\expandafter\let\csname equation*\endcsname\relax
\expandafter\let\csname endequation*\endcsname\relax
\usepackage{amssymb}
\usepackage{amsthm}
\usepackage{amstext}
\usepackage{mathtools}
\usepackage{graphicx}
\usepackage{placeins}
\usepackage{float}
\usepackage{subfigure}
\usepackage{CJK}
\usepackage{harvard}
\usepackage{calrsfs}

\begin{document}
\begin{CJK*}{GBK}{ }

%Unified Homogenization of Photonic/Phononic Crystals with First-band Negative Refraction 

\title[Sia Nemat-Nasser]{Unified Homogenization of Photonic/Phononic Crystals: Controllable First-band Negative Refraction }
%Homogenized Properties of Phononic and Photonic Crystals for Anti-plane Shear and Transverse Electric or Magnetic Waves: a Unified Approach

\author{Sia Nemat-Nasser}

\address{Department of Mechanical and Aerospace Engineering\\
University of California, San Diego\\
La Jolla, CA, 92093-0416 USA}
\ead{sia@ucsd.edu}
\vspace{10pt}
\begin{indented}
\item[February 27, 2016]
\end{indented}
\end{CJK*}

\begin{abstract}

It is shown, for the first time,  that negative refraction with positive phase velocity refraction can be realized (and controlled) over a wide range of frequency on the first (lowest) pass band of simple photonic and phononic crystals. 
 
First a unified approach is presented to accurately, efficiently, and uniquely produce the  homogenized effective material properties of doubly periodic phononic crystals for anti-plane shear (SH) and photonic crystals for transverse electric (TE) and transverse magnetic (TM) electromagnetic Bloch-form waves in such a manner that they exactly reproduce the band structure of the composite over any desired frequency band. 
Then the correspondence between phononic and photonic field equations is established and their effective homogenized material parameters are calculated.   
Finally illustrative examples are worked out for each case, revealing a rich body of refractive characteristics of these crystals, and showing by means of these examples that the homogenized effective parameters do yield the exact results, and that \textit{the resulting homogenized medium does in fact embody exactly the actual band structure and dispersive properties of the considered phononic/photonic crystal.}    
As a consequence when the homogenized medium is placed in contact with a normal homogeneous half-space, \textit{ it could display, even on the first pass band,  positive, negative, or even no energy refraction depending on
the frequency and wave vector of plane waves incident from the normal homogeneous solid to the interface.}

The unit cell of the crystal may consist of anisotropic constituents of any constant or variable properties that may also admit large discontinuities. 
The actual calculation procedure is detailed in \ref{AppendixA}, where,
for elliptical and rectangular inclusion geometries, analytical expressions are also given to assist the calculation of the key quantities.

\end{abstract}

\textit{Keywords}: First-band negative refraction; Effective  dynamic properties; Doubly periodic phononic and photonic composites; Band structure, group and energy-flux vectors

\section{Introduction}

A representative volume element  (RVE) of any heterogeneous material may be viewed as a unit cell of a periodic composite.  But photonic and phononic crystals with their periodic arrangement of identical unit cells of structured  constituent geometry and material properties, uniquely posses a wealth of special elastodynamic/electromagnetic characteristics that are not shared by an arbitrary RVE. These crystals display band structure and have dispersion responses that vary with frequency and wave vector,
\cite{yablonovitch1991photonic},
\cite{plihal1991photonic},
\cite{kushwaha1993acoustic},
\cite{pendry1994photonic}, 
\cite{sukhovich2008negative},
\cite{nemat2015anti},
\cite{nemat2015refraction}.
The overall response of these crystals is fully defined by the geometry and material properties of their typical unit cell.  Therefore, their homogenized effective properties for Bloch-form waves \cite{bloch1928quantum} can be calculated by simple spatial volume averaging of the field variables over a typical unit cell in such a manner that both continuity and periodicity are preserved. 
This approach was  introduced in \cite{willis2009exact}, \cite{PRBnemat2011} for one-dimensional phononic crystals, and it is the approach we use in the present work. 
The resulting constitutive material parameters, obtained by this averaging scheme, together with the corresponding averaged field equations, will then by necessity 
produce the exact band structure of the original heterogeneous crystal. They are functions of the wave frequency and wave vector, are unique, and fully embody the refractive characteristics of the original composite.
They may take on positive or negative values depending on the frequency and the wave vector.  Hence the homogenized material may behave either as a normal composite or it may behave as a so-called metamaterial.
We also show how this structure can be controlled and modified using  anisotropic matrix materials and simple unit-cell geometries.
We believe our results and the proposed method of analysis are new and should provide powerful tools for designing novel photonic or phononic crystals.
Indeed, exploiting this,
we show for the first time, that negative refraction with positive phase velocity refraction can be realized \textit{and controlled on a wide frequency-range  of the first (lowest) pass band} of simple photonic and phononic crystals. 

Luo et al. (2002) have demonstrated negative refraction in a photonic crystal consisting of circular holes placed in a square-lattice pattern within a homogeneous and isotropic dielectric matrix.  By interfacing the crystal with air along its (11)-direction, they observed that negative refraction for TE waves can occur in a narrow frequency (about 6.1\%) at the far edges of the first pass band.
In the present paper, we demonstrate a rich body of refractive characteristics that can be realized in photonic and phononic crystal over a wide region of the first pass band by suitably combining the square geometric anisotropy with material anisotropy, suggesting that this can be further enhanced by using a rectangular lattice structure.

An alternative method of homogenization is to consider a uniform reference medium with assumed constant properties and then introduce and calculate the necessary polarization strain-momentum or stress-velocity fields in the phononic case and polarization electric and magnetic fields in the photonic case such as to produce point-wise  the actual field variables.  For elastodynamic and electromagnetic homogenization using this approach, Willis has developed general variational principles \cite{willis1981variationalb}, \cite{willis1981variationala},  
\cite{willis1984variational}
\cite{milton2007modifications}.   The polarization approach has also been used by \cite{amirkhizi2008numerical} to homogenize the electromagnetic properties of periodic media (photonic crystals), and by the present author and Srivastava \cite{nemat2011overall}, \cite{srivastava2012overall}  to homogenize phononic crystals; see also
\cite{srivastava2015elastic} 
for comments. 

There are other  methods for calculating the effective properties; see for example,
\cite{joannopoulos2011photonic},
\cite{antonakakis2013asymptotics},
\cite{antonakakis2013high},
\cite{yang2014homogenization},
\cite{antonakakis2014homogenisation},
\cite{yang2014homogenization},
\cite{zhang2015effective},
and references cited therein.
Here we use the direct field averaging for calculation of the effective properties that requires expressions of the periodic part of the field variables.

%&&&&&&&&&&&&&&&&&&&&&&&&&&&&&&&&&&&&&&&&&&&&&&&&&&&&&&&&

The accurate calculation of these periodic parts for two- and three-dimensional phononic/photonic crystals is a formidable task for unit cells 
consisting of constituents of diverse properties with discontinuities. 
Naturally, the accuracy of the final results, whether obtained by direct calculation or through the corresponding effective
properties, will depend on the accuracy of the estimate of the periodic part of the field variables,
though both methods will yield exactly the same results.

There are only  limited computational tools that can be employed, each with its own limitations,
\cite{leung1990full},
\cite{bell1995program},
\cite{tran1995photonic},
\cite{pendry1996calculating},
\cite{birks2006approximate},
\cite{hussein2009reduced},
\cite{nemat2015refraction}.
%%%%%%%%%%%%% 
In the present work, we use a mixed variational formulation to calculate the band structure and the periodic part of the field variables \cite {NN-1972a}.
In this approach, both the displacement and stress fields in the phononic and the electric and magnetic fields in the photonic crystals are varied independently.  Hence they may be approximated by any  continuously differentiable  set of complete base functions, even though in actuality the gradients of some of the field variables may suffer large discontinuities across interfaces of various constituents of a typical unit cell. 
Being based on a variational principle, any set of approximating functions can be used for the calculations, e.g., plane-wave Fourier series or finite elements \cite{nemat1975harmonic}, 
\cite{minagawa1976harmonic}, \cite{minagawa1981finite}. 
The mix variational approach produces very accurate results and the rate of convergence of the corresponding series solution is superior to alternative methods \cite{babuska1978}. The general approach has been used to develop a fast and accurate computational platform for band structure calculations \cite{lu2016variational}, where it is shown that this approach provides a greater convergence rate than the usual Rayleigh quotient, especially in the presence of large material discontinuities; see also \cite{nemat2015anti} and \cite{nemat2015refraction}.  

%&&&&&&&&&&&&&&&&&&&&&&&&&&&&&&&&&&&&&&&&&&&&&&&&&&&&&&&&

\section{Statement of the Problem and Field Equations}

Consider a doubly periodic composite composed of rectangular unit cells of common dimensions $a_1$ and $a_2$. A typical unit cell, $\Omega_1$, includes a nested set of concentric rectangular or elliptical (or their combination) shaped inclusions, $\Omega_l$, $l=2,3,...,n$,
$ \Omega_1\supset \Omega_2 \supset \Omega_3 ~  {...}  \supset \Omega_n.$ 
For simplicity let the principal axes of the inclusions,
$a_1(j)$ and $a_2(j)$,  be parallel to the coordinate axes, 
${x}_1$ and ${x}_2$, although this is not necessary.
In what follows, we set $a_1=a_1(1)$ and $a_2=a_2(1)$.

We first consider the basic field equations and establish the one-to-one correspondence  between anti-plane shear  (SH) waves in phononic crystals and transverse electric (TE) and transverse magnetic  (TM) electromagnetic (EM) waves  in the photonic crystals.  Then we present a general method to produce the homogenized constitutive parameters for the SH Bloch-waves by averaging the periodic part of the field variables and, based on the correspondence among the field variables, we summarize the results for the TE and  TM, EM-waves.
The calculation details are given in section (\ref{AppendixA}).  In section (\ref{Examples}) we work out three numerical examples and also show that the homogenized effective medium does indeed embody the exact band structure of the original heterogeneous periodic composite,
which clearly distinguishes it from any anisotropic normal material. We also show that negative refraction can occur on the first (acoustic) pass band.

\subsection{Basic Field Equations: SH Waves} 

The Bloch-form SH waves of frequency $\omega$ involve two shear stresses,
 $\tau_j(x_1,x_2)$, $j=1,2$, and a lateral displacement,
 $w(x_1,x_2)$, satisfying the following field equations:
\begin{eqnarray}\label{SH1} %1
	  \begin{array}{c}
 	 \tau_{j,j}+\omega^2 \rho w=0,\quad
	 \gamma_{j}=w_{,j},\quad\\
	 \tau_j=\mu_{jk}\gamma_k,\quad
	  \gamma_{j}=D_{jk}\tau_k,\quad
	  j,k=1,2.
	  \end{array}
\end{eqnarray}
Here $\rho$ is the mass density, $\mu_{jk}=\mu_{kj}$ and $D_{jk}=D_{kj}$ are the components of the shear and compliance moduli, respectively, repeated indices are summed, and a comma followed by an index denotes differentiation with respect to the corresponding coordinate variable. When the principal axes of the elasticity tensor, $\mu_{jk}$, are in the $x_1, x_2$-directions, then $\mu_{12}=\mu_{21}=0$. 
In what follows we consider the most general case where
$\mu_{12}=\mu_{21}\neq 0$ and $\mu_{11}\neq\mu_{22}$, and 
show by means of illustrative examples how the resulting band structure is affected by these factors. 

\subsection{Basic Field Equations: TE/TM Waves}

Consider now the basic equations for EM waves and denote by 
$\mathcal{E}_j$, $\mathcal{H}_j$, $\mathcal{B}_j$, and  $\mathcal{D}_j$, $j=1,2,3$, 
the components of  electric,  magnetic, magnetic induction, and electric displacement fields, respectively.
Here we focus on TE and TM electromagnetic waves, set 
$\mathcal{E}_3=\mathcal{H}_1=\mathcal{H}_2=0$ and $\mathcal{H}_3=\mathcal{H}$ for TE waves, and
$\mathcal{H}_3=\mathcal{E}_1=\mathcal{E}_2=0$ and $\mathcal{E}_3=\mathcal{E}$ for TM
waves, reducing the general Maxwell equations, 
\begin{equation}\label{GEM1} %2
\nabla\times\mathcal{E}-i\omega\mathcal{B}=0,~~ 
\nabla\times\mathcal{H}+i\omega\mathcal{D}=0,
\end{equation}
\begin{equation}\label{GEM2} %3
\nabla . ~\mathcal{B}=\nabla . ~\mathcal{D}=0,~~
\mathcal{D}_j=\epsilon_{jk}\mathcal{E}_k,~~
\mathcal{B}_j=\mu_{jk}\mathcal{H}_k,
\end{equation}
to the following set of scalar equations, for TE and TM waves, respectively:
\begin{eqnarray}\label{TEEM1} %4
\begin{array}{c}
\mathcal{E}_{1,2}-\mathcal{E}_{2,1}+i\omega {\mu}\mathcal{H}=0,~~~\quad\\
\mathcal{H}_{,2}+i\omega (\epsilon_{11} \mathcal{E}_1+
\epsilon_{12} \mathcal{E}_2)=0,\quad\\
\mathcal{H}_{,1} -i\omega (\epsilon_{21} \mathcal{E}_1+\epsilon_{22} \mathcal{E}_2)=0~~~,
\end{array}
\end{eqnarray}
\begin{eqnarray}\label{TMEM1}%5
\begin{array}{c} 
-\mathcal{H}_{1,2}+\mathcal{H}_{2,1}+i\omega\epsilon\mathcal{E}=0,~~~~\quad\\
\mathcal{E}_{,2}-i\omega( \mu_{11} \mathcal{H}_1+
\mu_{12}\mathcal{H}_2)=0,\quad\\
\mathcal{E}_{,1} +i\omega( \mu_{21} \mathcal{H}_1+
 \mu_{22} \mathcal{H}_2)=0~~~.
\end{array}
\end{eqnarray}
In equations (\ref{TEEM1}), 
$\mu= \mu_{33}$ is the magnetic permeability and $\epsilon_{jk}=\epsilon_{kj}$ are the components of the electric permittivity, and in equations (\ref{TMEM1}),
$\epsilon= \epsilon_{33}$ is the electric permittivity and $\mu_{jk}=\mu_{kj}$ are the components of the magnetic permeability, respectively.  All the field variables are functions of 
$x_1, x_2$ only since the time variable is incorporated into the field equations.
When the principal axes of the material parameters are, respectively, in the 
$x_1,x_2$-directions, then $\epsilon_{12}=\epsilon_{21}=0$ and
$\mu_{12}=\mu_{21}=0$. Here again we do not assume this and consider the most general anisotropic case.

\subsection{Correspondence between SH and TE/TM Waves}

The field equations (\ref{TEEM1})  reduce to field equations (\ref{SH1}) if we set,
\begin{equation}\label{SHTE}%6
\mathcal{E}_{1}=\tau_2,\quad
\mathcal{E}_{2}=-\tau_1,\quad
\mathcal{H}=-i\omega w,\quad
\mu=\rho,\quad
\epsilon_{jk}=D_{jk}.
\end{equation}
Similarly, the field equations (\ref{TMEM1}) reduce to field equations (\ref{SH1}) if we set,
\begin{equation}\label{SHTM}%7
\mathcal{H}_{1}=-\tau_2,\quad
\mathcal{H}_{2}=\tau_1,\quad
\mathcal{E}=-i\omega w,\quad
\epsilon=\rho,\quad
\mu_{jk}=D_{jk}.
\end{equation}
Therefore, it is sufficient to outline the homogenization procedures for the SH stress waves and then extract the corresponding homogenized results for TE and TM waves by direct association based on equations (\ref{SHTE}) and (\ref{SHTM}).

To this end we average the periodic parts of the field equations as well as the
periodic parts of the constitutive equations over a typical unit cell of the composite.
In this manner the effective parameters are obtained independently of the conservation equations. Substitution of the resulting effective parameters into the averaged conservation relations should then yield the exact band structure in each case.  The method in universal and can be used in two- and three-dimensional cases with impunity. 
% &&&&&&&&&&&&&&&&&&&&&&&&&&&&&&&&&&&&&&&&&&&&&&&&&&&&&&&&&&&&&&&&&&&&&&&
% see Nemat-Nasser (2016) for application to two dimensional phononic crystals. 
% &&&&&&&&&&&&&&&&&&&&&&&&&&&&&&&&&&&&&&&&&&&&&&&&&&&&&&&&&&&&&&&&&&&&&&&

%
\section{Homogenization}
We now seek to extract the overall dynamic effective material parameters of  the composite such that they produce the exact frequency band structure of the composite.  

\subsection{Homogenized Effective Parameters: SH Waves}

For Bloch-form time-harmonic SH waves of frequency $\omega$,
traveling in a doubly periodic  phononic crystal,
the time-dependence of the field variables is incorporated into equations(\ref{SH1}).
These field variables have the following structure:
\begin{eqnarray}\label{SH2}%8
\left[ 
\begin{array}{c}
w\\
\gamma_j\\
\tau_j\\
\end{array} \right]=
\left[ \begin{array}{c}
w^p(x_1,x_2)\\
\gamma_j^p(x_1,x_2)\\
\tau_j^p(x_1,x_2)\\
\end{array} 
\right]e^{i(k_1x_1+k_2x_2)};\quad j=1,2,
\end{eqnarray}
where superimposed $p$ denotes the periodic part, and
$k_1$ and $k_2$ are the components of the wave vector. 

\subsubsection{Average Effective Constitutive Parameters:}  

We directly average the periodic parts of equations (\ref{SH2}) over a typical unit cell and set,
\begin{eqnarray}\label{HSH1}%9
\left[ 
\begin{array}{c}
\bar{w}\\
\bar{\gamma}_j\\
\bar{\tau}_j\\
\end{array} \right]=
\left[ \begin{array}{c}
\bar{w}^p\\
\bar{\gamma}_j^p\\
\bar{\tau}_j^p\\
\end{array} 
\right]e^{i(k_1x_1+k_2x_2)};\quad j=1,2,
\end{eqnarray}
where a typical averaged  
field variable, say
$\bar{G}$,
is defined as
\begin{equation}\label{HG}%10
\bar{G}=\bar{G}^p e^{i(k_1x_1+k_2x_2)},~~~~
\bar{G}^p=\int_{-a_1/2}^{a_1/2}\int_{-a_2/2}^{a_2/2} G^p({x_1,x_2})dx_1 x_2.
\end{equation}
We now define
the effective compliance and stiffness tensors, $D_{jk}^{eff}$ and $\mu_{jk}^{eff}$,  and the effective mass density, $\rho^{eff}$,  by
\begin{equation}\label{direct}%11
\overline{(D_{jk}\tau_k^p)}=D_{jk}^{eff}\bar{\tau}_{k}^p,\qquad
\mu_{jk}^{eff}=[D_{jk}^{eff}]^{-1},\qquad
\overline{(\rho w^p)}=\rho^{eff}\bar{w}^p,
\end{equation}
where the quantities
$\overline{(D_{jk}\tau_k^p)}$ and $\overline{\rho w^p}$ are calculated by direct averaging over the unit cell.
Using the averaged field quantities, (\ref{direct}), the field equations (\ref{SH1}) are now averaged (see \ref{AppendixB} for details) to obtain, 
\begin{eqnarray}  \label{HSH2}%12
 \begin{aligned}
&ik_{1}\bar{\tau}_{1}+ik_{2}\bar{\tau}_{2}+\omega^2 {\rho}^{eff}\bar{ w}=0,~~
ik_j \bar{w}-\bar{\gamma}_j=0,~~\\
&\bar{\gamma}_j-D^{eff}_{jk}\bar{\tau}_k=0,~~
\bar{\tau}_j-\mu^{eff}_{jk}\bar{\gamma}_k=0,~~
[\mu^{eff}_{jk}]=[D^{eff}_{jk}]^{-1}.
\end{aligned}
\end{eqnarray} 
The band structure is obtained from equations (\ref{HSH2}) by substituting  
(\ref{HSH2})$_{2,3,4}$ into (\ref{HSH2})$_1$, arriving at,
\begin{equation}\label{HSH3}%13
{\omega^{2}}=\frac{1}{\rho^{eff}}
k_{j}\mu_{jl}^{eff}k_l,\quad
j,l=1,2~~~
for~SH,~ stress~waves,
\end{equation}
where the repeated indices are summed. Note that the effective compliance tensor,
 $D_{jk}^{eff}$, and hence the corresponding stiffness tensor, $\mu_{jk}^{eff}$,  as well as the effective mass density, $\rho^{eff}$, are calculated directly once the periodic parts of the stress and displacement fields are known. Although these fields are calculated independently using a mixed variational method, 
as discussed in \ref{AppendixA}, they are the actual field quantities that correspond to the exact associated frequency band.  Therefore they should, by necessity, produce the correct band structure.
It is however important to note that \textit{the equation of motion}, (\ref{HSH2})$_1$, \textit{is not, and should not be used to obtain the effective constitutive parameters}.
These effective parameters should characterize the considered particular material (here a linearly elastic one), independently of the conservation laws (here the linear momentum).  The author has used a similar direct averaging technique to calculate the effective properties of phononic crystals in plane strain and plane stress cases where there are two coupled conservation equations. The procedure does indeed yield the exact band structure and the resulting effective compliance components do automatically satisfy the stress compatibility equation.

% $$$$$$$$$$$$$$$$$$$$$$$$$$$$$$$$$$$$$$$$$$$$

\subsection{Homogenized Effective Parameters: TE/TM Waves}\label{junk}

For the TE Bloch-form waves, set
\begin{eqnarray}\label{HTE1}%14
\left[ \begin{array}{c}
\mathcal{H}\\
\mathcal{E}_j\\
\end{array} \right]=
\left[ \begin{array}{c}
\mathcal{H}^{p}\\
\mathcal{E}_j^{p}\\
\end{array} \right]e^{i(k_1x_1+k_2x_2)},
\end{eqnarray}
and for the TM Bloch-form waves, set 
\begin{eqnarray}\label{HTM1}%15
\left[ \begin{array}{c}
\mathcal{E}\\
\mathcal{H}_j\\
\end{array} \right]=
\left[ \begin{array}{c}
\mathcal{E}^{p}\\
\mathcal{H}_j^{p}\\
\end{array} \right]e^{i(k_1x_1+k_2x_2)}.
\end{eqnarray}

%%%%%%%%%%%%%
Now consider equations (\ref{TEEM1},\ref{TMEM1}), and following similar  steps as before, obtain the following homogenized TE and TM wave equations, respectively:
\begin{eqnarray}  \label{HTE2}%16
\begin{aligned}
&k_2\bar{\mathcal{E}}_1-k_1\bar{\mathcal{E}}_2+
\omega \mu^{eff} \bar{\mathcal{H}}=0,~~\\
&k_1\bar{\mathcal{H}}-\omega({\epsilon}^{eff}_{21}\bar{\mathcal{E}}_1+
{\epsilon}^{eff}_{22}\bar{\mathcal{E}}_2)=0,~~\\
&k_2\bar{\mathcal{H}}+\omega 
({\epsilon}^{eff}_{11}\bar{\mathcal{E}}_1+
{\epsilon}^{eff}_{12}\bar{\mathcal{E}}_2)=0;
\end{aligned}
\end{eqnarray} 
\begin{eqnarray}\label{HTM2}%17
\begin{aligned}
&k_1\bar{\mathcal{H}}_2-k_2\bar{\mathcal{H}}_1+
\omega \epsilon^{eff}\bar{\mathcal{E}}=0,~~\\
&k_2\bar{\mathcal{E}}-
\omega( {\mu}^{eff}_{11}{\mathcal{H}}_1+
{\mu}^{eff}_{12}{\mathcal{H}}_2)=0,~~\\
&k_1\bar{\mathcal{E}}+
\omega ({\mu}^{eff}_{21}\bar{\mathcal{H}}_1+
{\mu}^{eff}_{22}\bar{\mathcal{H}}_2)=0,
\end{aligned}
\end{eqnarray} 
where superimposed bar denotes the average value of the corresponding quantity, taken over a  unit cell, and
the effective properties for the TE and TM waves are defined such that,
\begin{equation}\label{EFFECTIVE3}%18
\overline{(\mu\mathcal{H}^p)}=\mu^{eff}\bar{\mathcal{H}^p},\quad
\overline{(\epsilon_{jk}\mathcal{E}^p_k)}= 
{\epsilon}^{eff}_{jk}\bar{\mathcal{E}}^p_k,\quad
[\nu^{eff}_{jk}]=[\epsilon^{eff}_{jk}]^{-1},\quad
for~TE,~EM~waves;
\end{equation}
\begin{equation}\label{EFFECTIVE4}%19
\overline{(\epsilon\mathcal{E}^p)}=\epsilon^{eff}\bar{\mathcal{E}^p},\quad
\overline{(\mu_{jk}\mathcal{H}^p_k)}= 
{\mu}^{eff}_{jk}\bar{\mathcal{H}^p_k},\quad
[\lambda^{eff}_{jk}]=[\mu^{eff}_{jk}]^{-1},\quad
for~TM,~EM~waves.
\end{equation}
To express the band structure of the photonic crystal in terms of the corresponding effective properties, we 
combine equations (\ref{HTE2} to \ref{EFFECTIVE4}) and obtain,
\begin{equation}\label{HTE4}%20
\omega^2=\frac{1}{{\mu}^{eff}}
(k^2_2\nu^{eff}_{11}+k^2_1\nu^{eff}_{22}-
k_2k_1(\nu^{eff}_{12}+\nu^{eff}_{21})),
~~~for~TE,~EM~waves;
\end{equation}
\begin{equation}\label{HTM4}%21
\omega^2=\frac{1}{{\epsilon}^{eff}}
(k^2_2\lambda^{eff}_{11}+k^2_1\lambda^{eff}_{22}-
k_2k_1(\lambda^{eff}_{12}+\lambda^{eff}_{21})),~~~for~TM,~EM~waves.
\end{equation}

For numerical examples, we first calculate the field variables associated with each frequency band, and then, follow the procedure outlined above, appropriately average over a unit cell to obtain the effective parameters.

%&&&&&&&&&&&&&&&&
%$$$$$$$$$$$$$$$$$$$$

\section{Illustrative Examples}\label{Examples}

We consider a simple two-phase photonic crystal that consists of a homogeneous anisotropic dielectric matrix containing 1.5mm holes, periodically spaced 4mm apart in both  the $x_1$- and   $x_2$-directions.
 Using thermoplastic co-extrusion fabrication or other methods, low-loss dielectric composites of various degrees of anisotropy can be fabricated; see, e.g., \cite{zhang2008fabrication}, \cite{wing2006fabrication}. 
 Here, for illustration, we use a matrix with dielectric constants  
of $\epsilon_{I}^{M}= 9.6$ and
$\epsilon_{II}^{M}= 90$, where the superimposed \textit{M} denotes the matrix dielectric tensor.

To examine the effect of the microstructure of the unit cell on the response of the resulting medium, we consider two cases that differ from one another only by a different arrangement of the principal axes of the matrix dielectric tensors relative to the $x_1, x_2$-coordinate directions, i.e., in relation to the periodicity of the original composite.  These cases are: 
\begin{description}
\item  \textit{Case A}: The principal axes of $\epsilon_{jk}^{M}$ coincide with the $x_1, x_2$-axes;
\item   \textit{Case B}: The principal axes of $\epsilon_{jk}^{M}$ make a 15$^o$ angle with the $x_1, x_2$-axes.
\end{description}
We then show
how the refractive behavior of the composite has been dramatically changed
by this simple rearrangement.  Indeed, \textit{we show, for the first time, that even on the first (lowest) pass band of the medium, one can realize negative refraction over a broad range of frequencies and phase angles.} In addition, we show that the frequencies at which such response occurs can be lowered by surrounding the central hole with 0.5mm thick high-dielectric material; such high dielectric solids can be produced by various means, see, e.g., \cite{sengupta1999breakthrough}, \cite{si2002epitaxial}.

In addition to the above two- and three-phase \textit{photonic} composites, we also consider a very simple \textit{phononic} crystal to illustrate our general results and show that negative energy refraction with positive phase-velocity refraction can be realized on the first (lowest) pass band. The constituent properties of the phononic crystal are chosen to have large variations within the unit cell to also show the effectiveness of the variational approach. These examples show the rich body of physics that can be revealed by the present approach as well as  the versatility and effectiveness of the proposed computational tool; see also \cite{nemat2015refraction}.  

\subsection{Unit-cell Properties for Photonic Crystal: TE-waves}

\begin{figure}%f1
\centering
\begin{minipage}[b]{0.45\linewidth}
\includegraphics[scale=0.40, trim=1cm 6cm 0cm 5cm, clip=true]{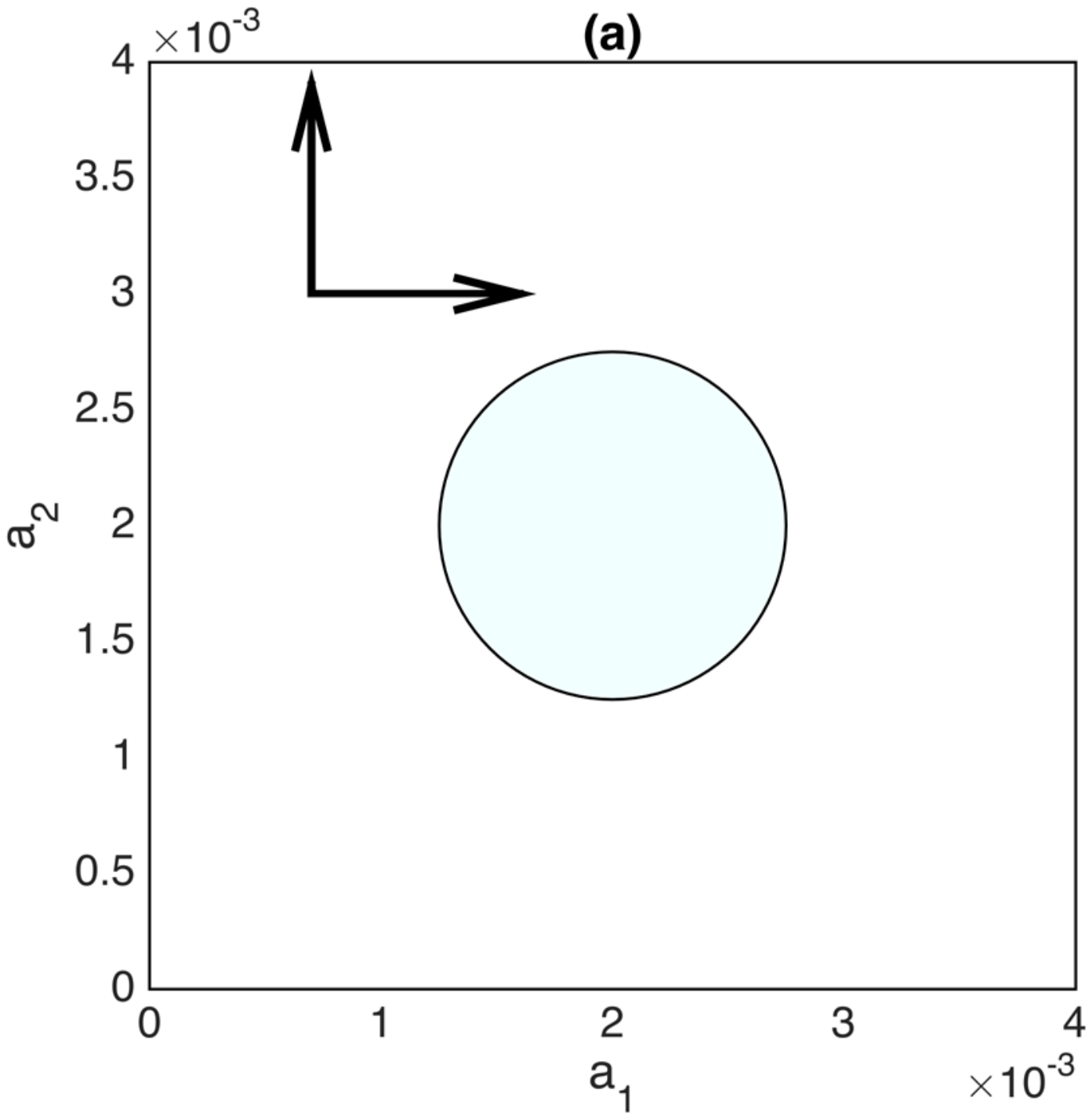}
\end{minipage}
\begin{minipage}[b]{0.45\linewidth}
\includegraphics[scale=0.40, trim=1cm 6cm 0cm 5cm, clip=true]{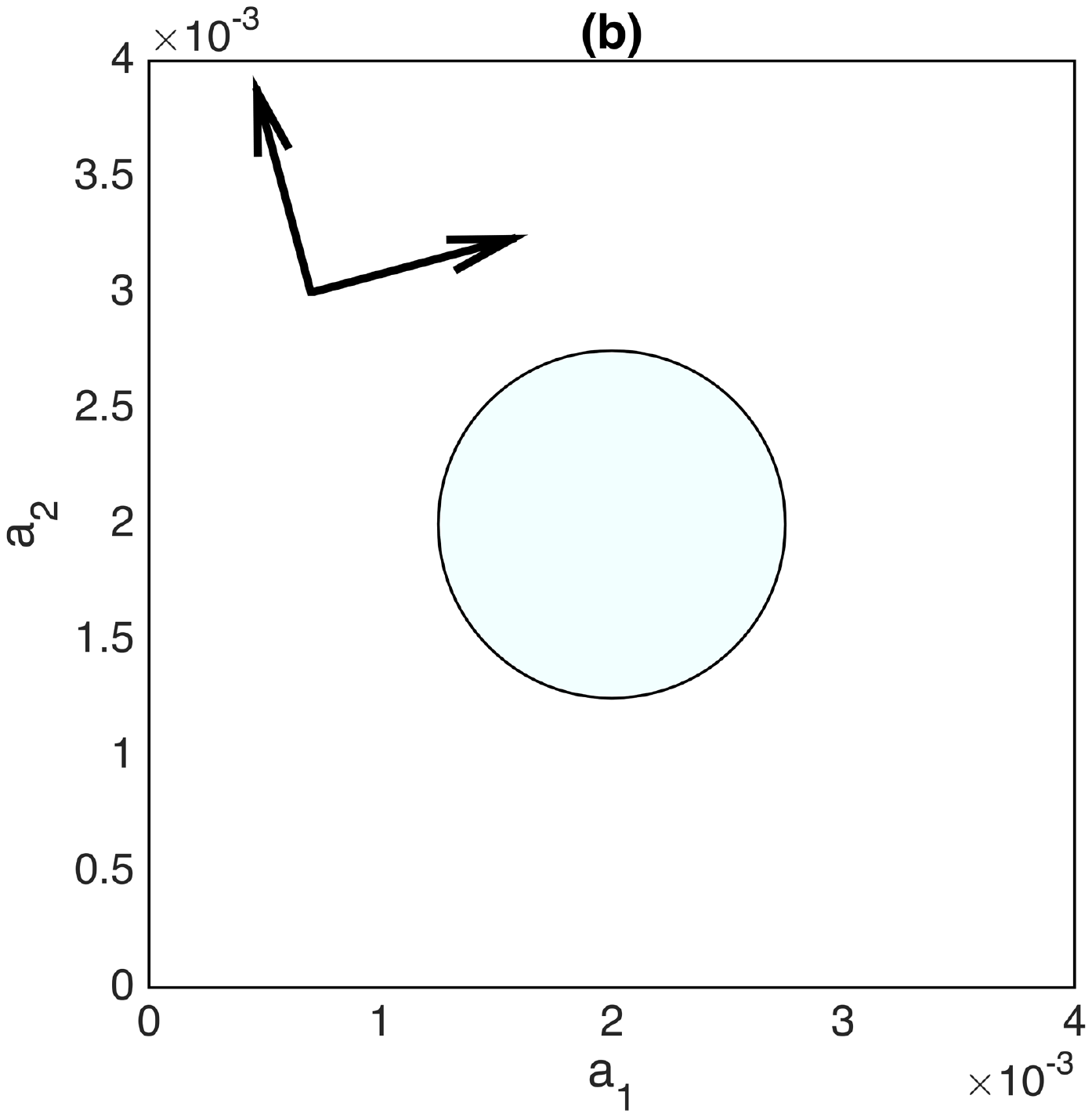}
\end{minipage}
\caption{The unit cell of a two-phase photonic crystal consisting of an anisotropic matrix with a central hole. (a): The principal axes (heavy arrows) of $\epsilon_{jk}^{M}$ coincide with the $x_1, x_2$-axes;
and (b): The principal axes (heavy arrows) of $\epsilon_{jk}^{M}$ make a 15$^o$ angle with the $x_1, x_2$-axes. }
\label{unit_cells}
\end{figure}

Figures  (\ref{unit_cells}) show the unit cells of the two photonic crystals that are examined below. The arrows on the upper left corner of each unit cell indicate the directions of the principal axes of the dielectric tensor of the matrix material. In Figure  (\ref{unit_cells}a) these principal axes coincide with the $x_1, x_2$-axes, whereas in Figure  (\ref{unit_cells}b) they are at a 15$^o$ angle with respect to the $x_1, x_2$-axes.  The normalized wave-vector components are
$Q_1=a_1k_1$ and $Q_2=a_2k_2$; note that $a_1= a_2=4mm$ in the present case.

The matrix electric constants in Figure  (\ref{unit_cells}a) are:
\begin{itemize}
\item $\epsilon_{11}^{M}=9.6$; 
\item $\epsilon_{22}^{M}=90$; 
\item $\epsilon_{12}^{M}=\epsilon_{12}^{M}=0$;
\item $\mu^{M} =1$.
\end{itemize}
The dielectric tensor  for Figure  (\ref{unit_cells}b), with the same principal values as in  Figure  (\ref{unit_cells}a), has the following components with respect to the $x_1, x_2$-coordinates:
\begin{itemize}
\item $\epsilon_{11}^{M}=15$; 
\item $\epsilon_{22}^{M}=84.6$; 
\item $\epsilon_{12}^{M}=\epsilon_{12}^{M}=20.1$,
\end{itemize}
with $\mu^{M}=1$.

\begin{figure}%\label{first10bands}%f2
\centering
\begin{minipage}[b]{0.45\linewidth}
\includegraphics[scale=0.45, trim=0cm 6cm 0cm 5cm, clip=true]{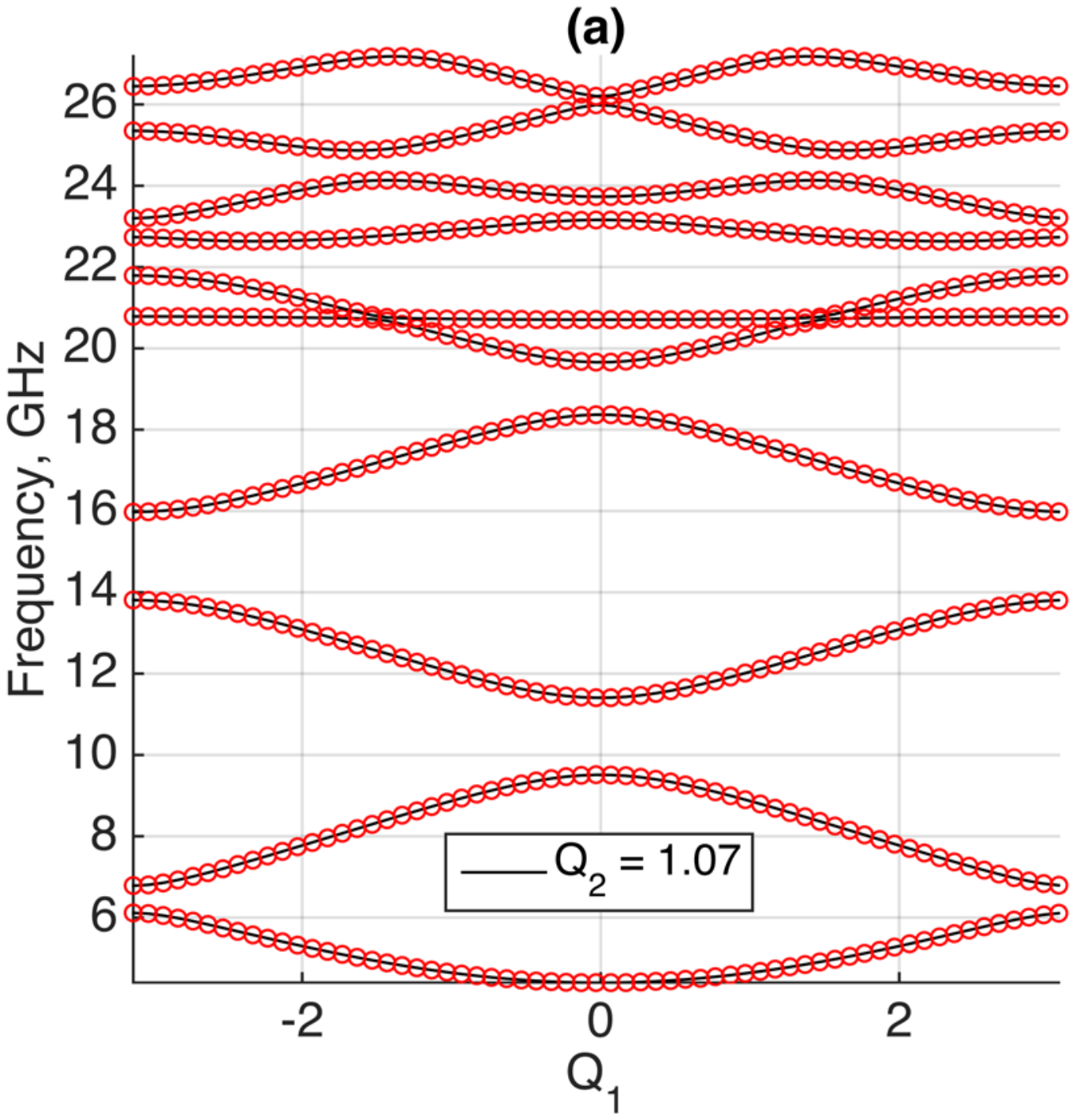}
\end{minipage}
\quad
\begin{minipage}[b]{0.45\linewidth}
\includegraphics[scale=0.45, trim=0cm 6cm 0cm 5cm, clip=true]{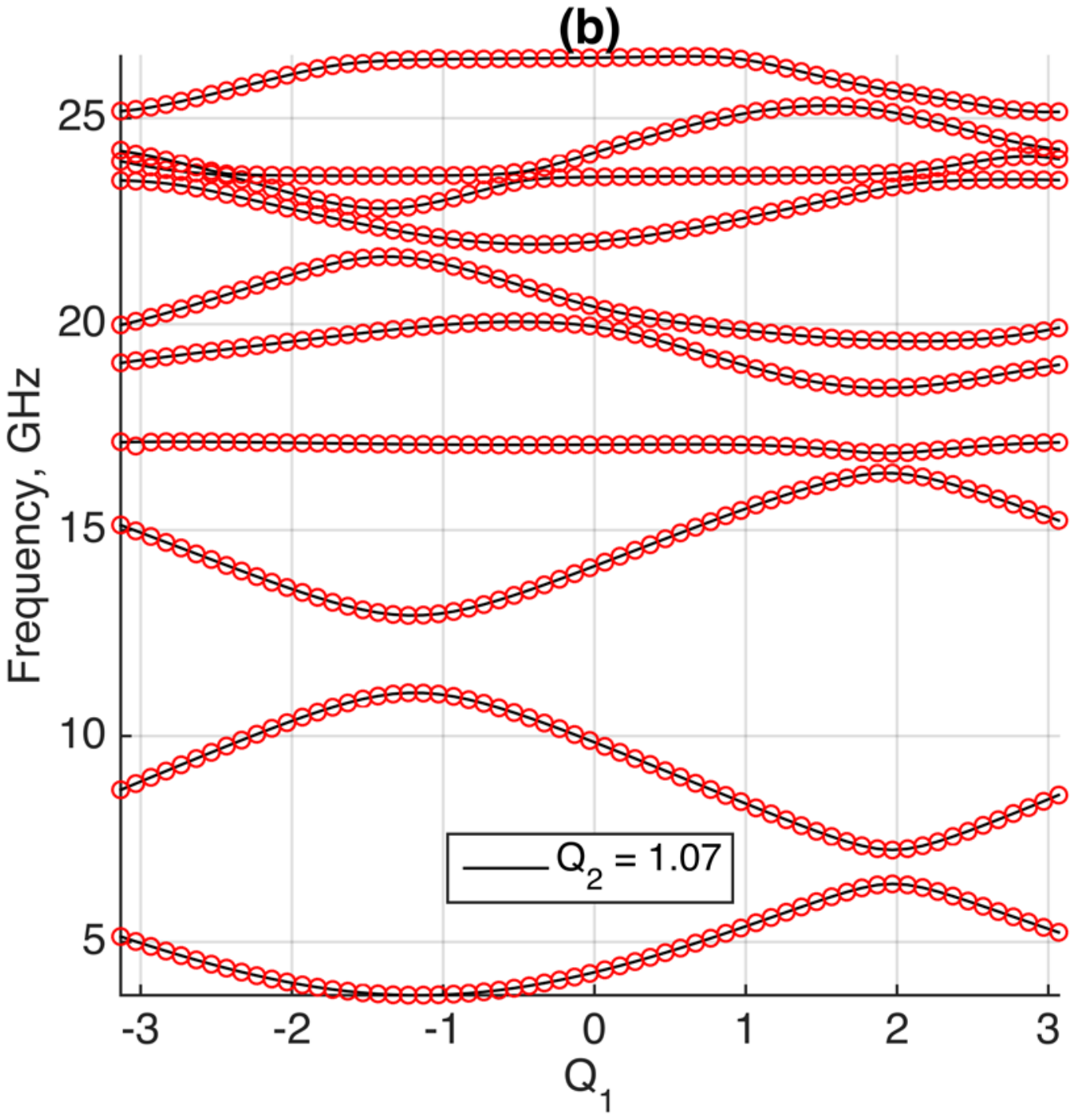}
\end{minipage}
\caption{The first ten pass bands for $Q_2=1.07$; solid lines are by direct calculation using equations (\ref{TEB1})$_1$; open circles are by  homogenized parameters using equation (\ref{HTE4}).  (a): Principal axes of electric permittivity tensor of matrix material  coincide; (b): they are at about 15$^o$ angle with respect to the $x_1,x_2$-axes; two-phase photonic crystal of unit cells  shown in Figures (\ref{unit_cells}a,b).}
\label{first10bands}
\end{figure}

The first 10 pass bands  corresponding to these unit cells are shown in
Figures \ref{first10bands} for $Q_2=1.07$.
In these figures, the solid lines are obtained by direct band structure calculation using equation (\ref{TEB1})$_1$ whereas the open circles are by using 
the effective parameters; equation (\ref{HTE4}). 
As can be seen, the effective parameters yield the  exact band structure in each case.
In what follows, the effective parameters are used for all related calculations.
%

%%%%%%%
\begin{figure}%\label{2phBs}%f3
\centering
\begin{minipage}[b]{0.45\linewidth}
\includegraphics[scale=0.42, trim=0cm 6.5cm 0cm 5cm, clip=true]{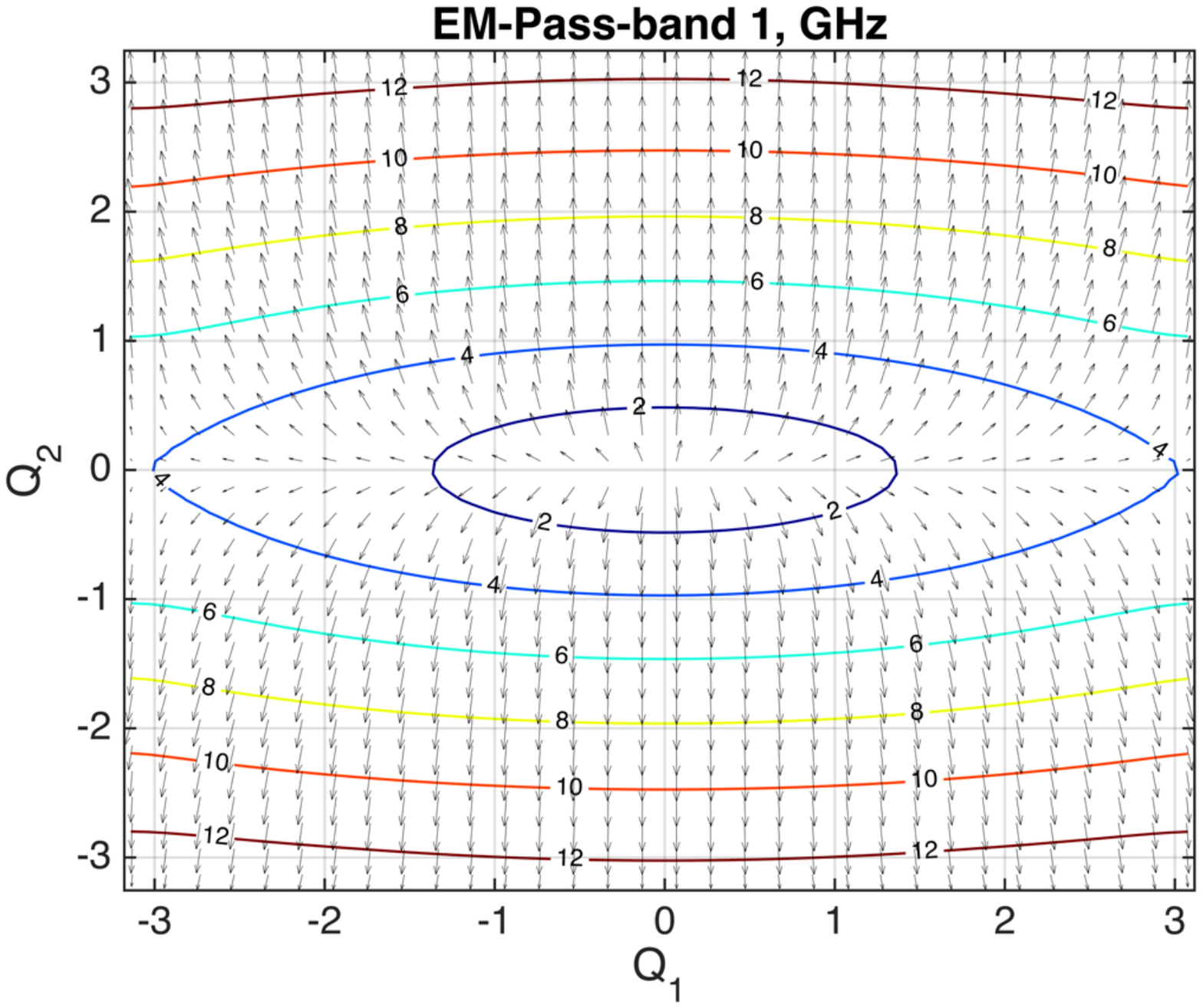}
\end{minipage}
\quad
\begin{minipage}[b]{0.45\linewidth}
\includegraphics[scale=0.42, trim=0cm 6.5cm 0cm 5cm, clip=true]{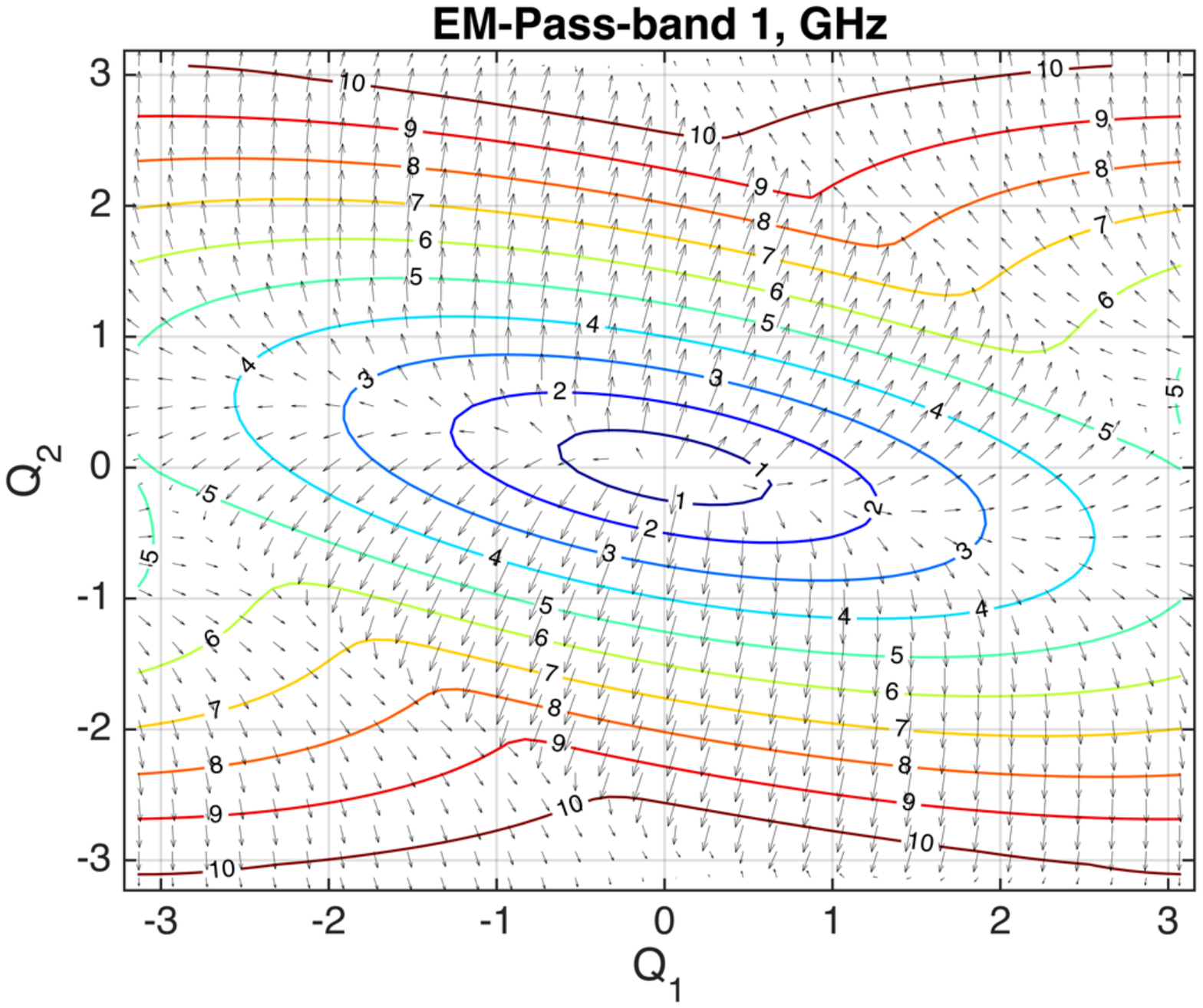}
\end{minipage}
%\quad
\begin{minipage}[b]{0.45\linewidth}
\includegraphics[scale=0.42, trim=0cm 6.5cm 0cm 5cm, clip=true]{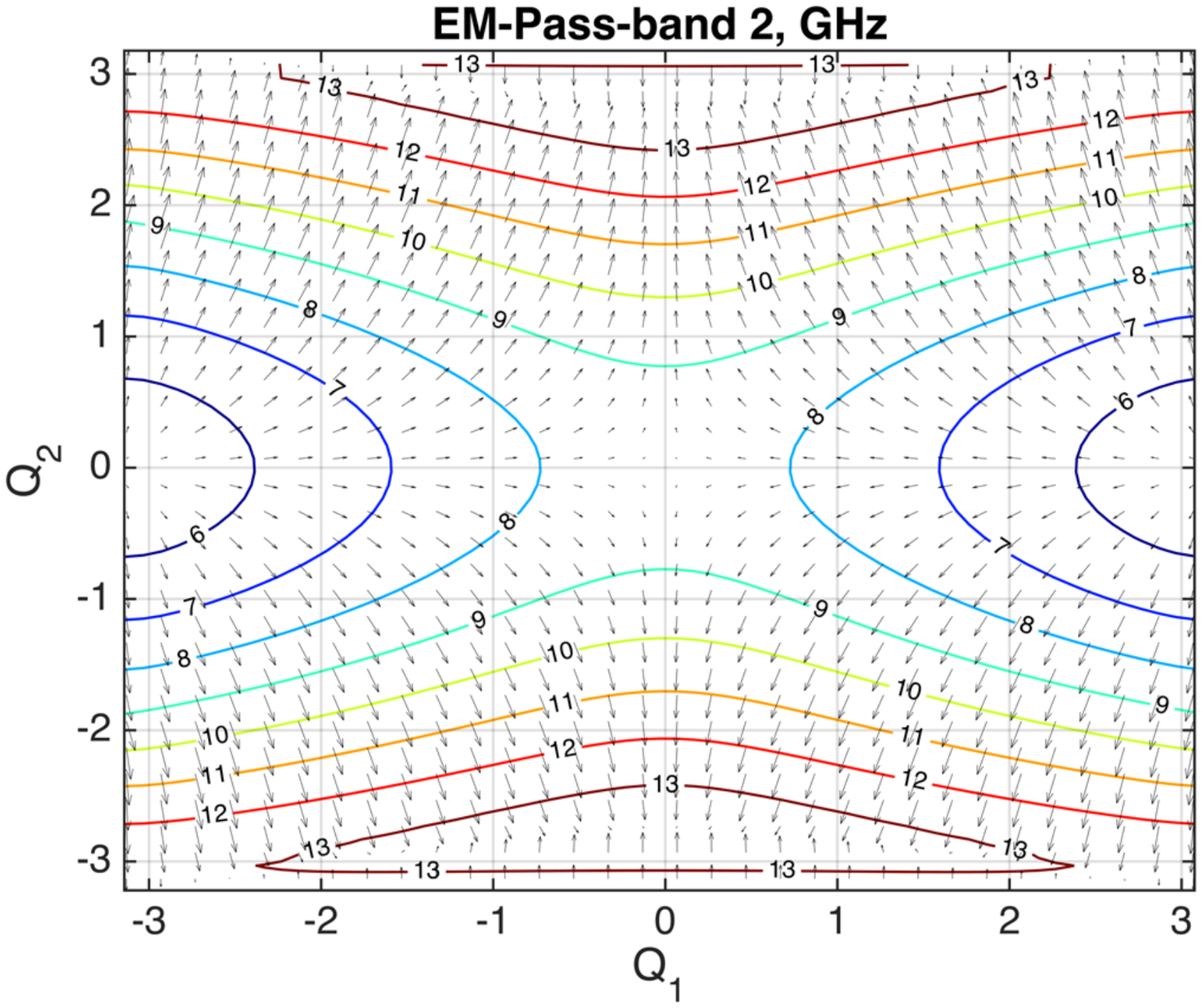}
\end{minipage}
\quad
\begin{minipage}[b]{0.45\linewidth}
\includegraphics[scale=0.42, trim=0cm 6.5cm 0cm 5cm, clip=true]{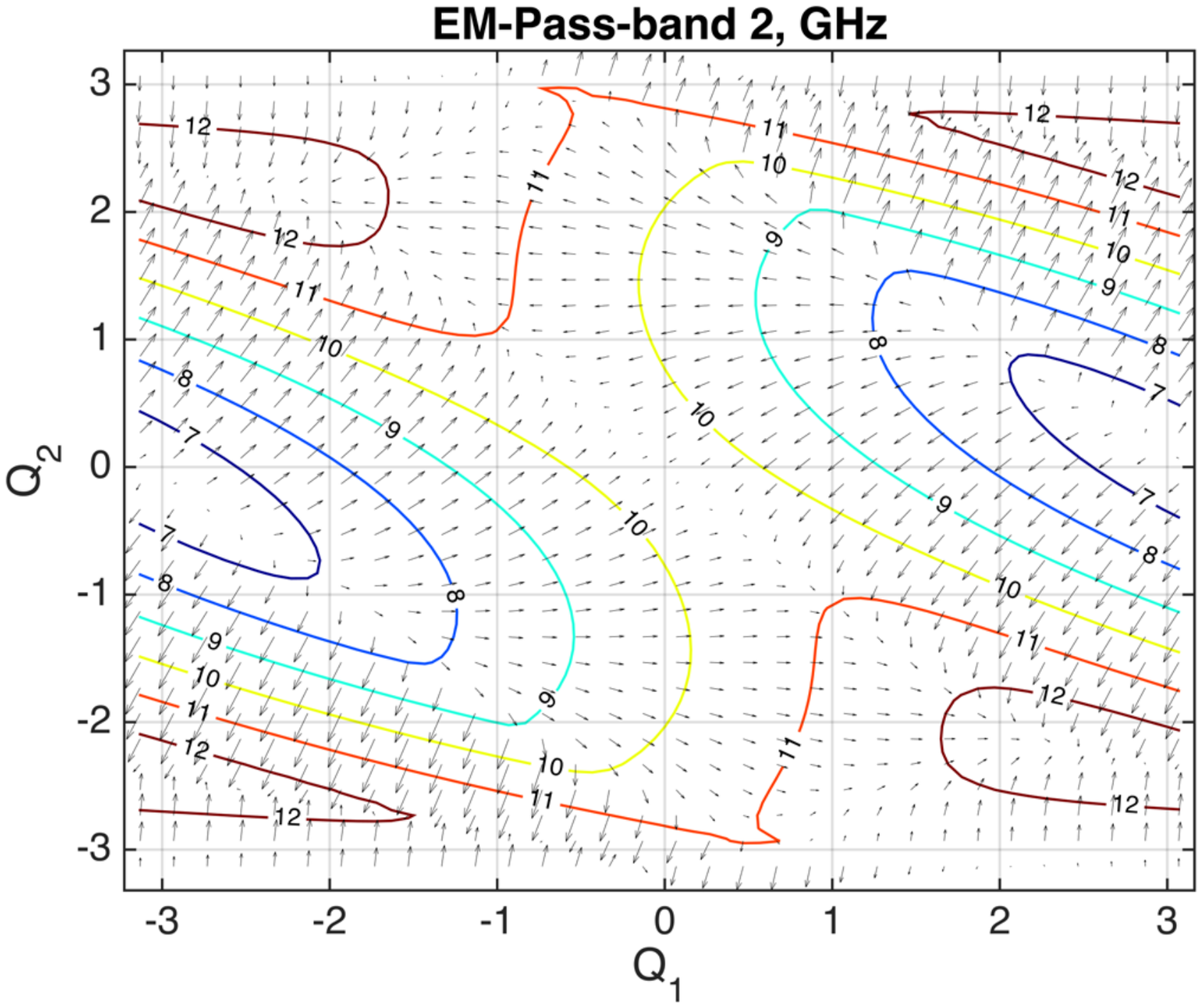}
\end{minipage}
%\quad
\begin{minipage}[b]{0.45\linewidth}
\includegraphics[scale=0.42, trim=0cm 6.5cm 0cm 5cm, clip=true]{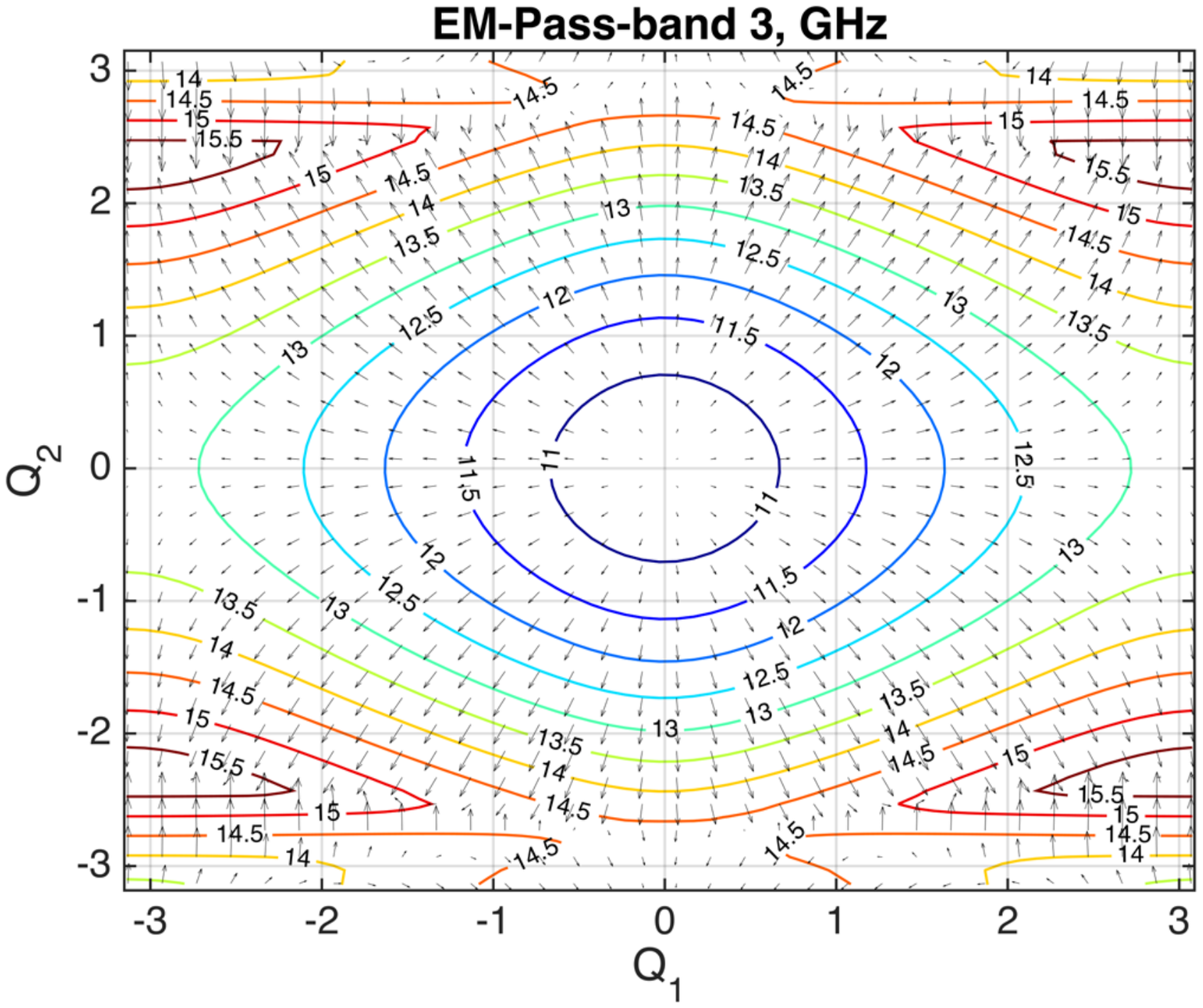}
\end{minipage}
%\quad
\begin{minipage}[b]{0.45\linewidth}
\includegraphics[scale=0.42, trim=0cm 6.5cm 0cm 5cm, clip=true]{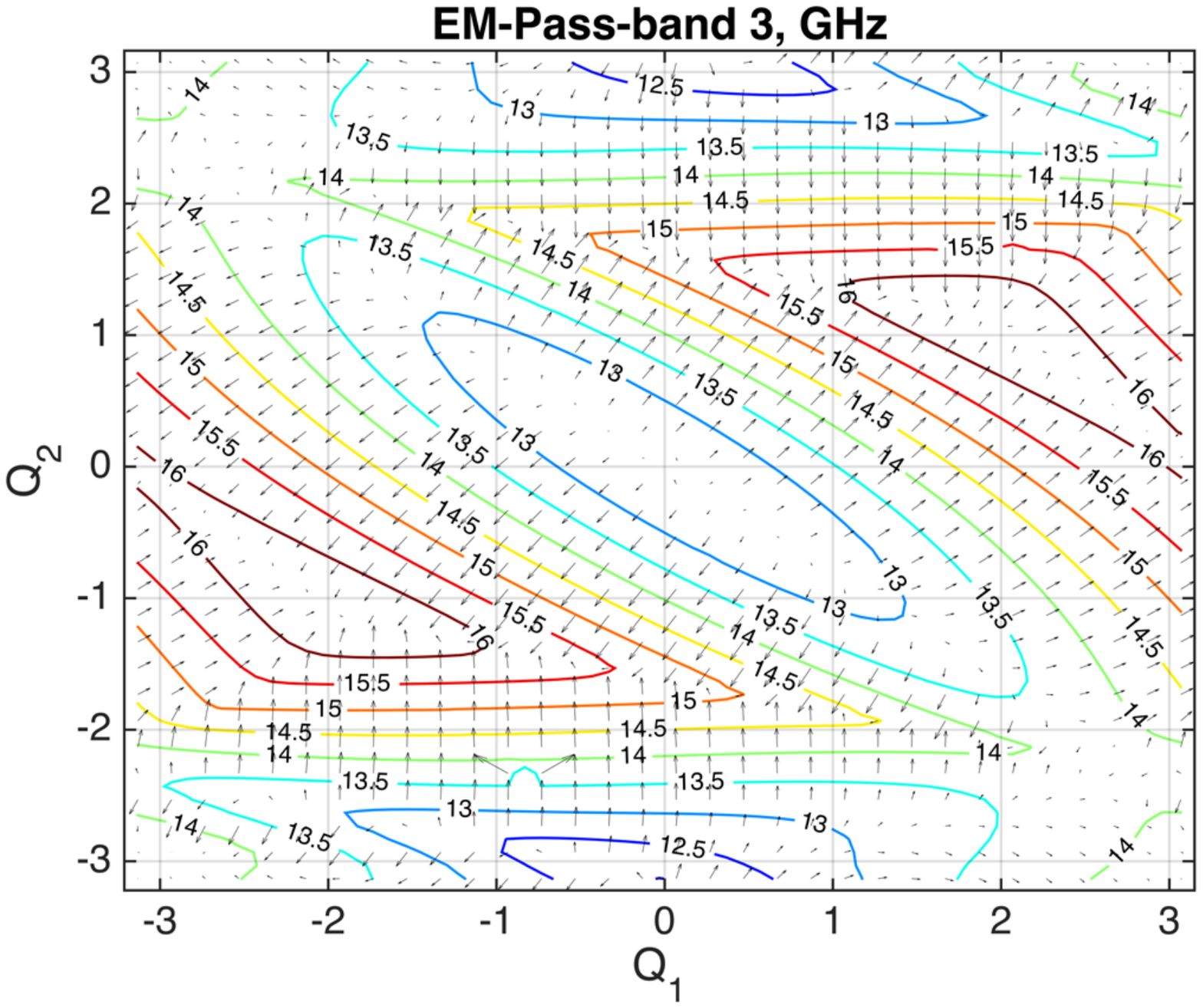}
\end{minipage}
\caption{Equi-frequency contours of the first three pass bands of the photonic crystals  
 Figures (\ref{unit_cells}a,b). Principal axes of electric permittivity tensor of matrix material  coincide (Left column), are at about 15$^o$ angle  (Right column) with the coordinate axes; arrows are the energy-flux vectors.}
\label{2phBs}
\end{figure}

In Figures (\ref{2phBs}) we have shown the equi-frequency contours of the first three pass bands
of these photonic crystals, calculated using the homogenized parameters.  Identical results are obtained from the direct calculations.  

The graphs in the left column in Figures  \ref{2phBs} are for the case when the principal axes of the electric permittivity tensor of the matrix material coincide with the geometric symmetry axes of the periodic composite, and those on the right are for the case when the electric permittivity and the geometric principal axes are at about 15$^o$ angle relative to one another.

The energy-flux vectors are also shown for these pass bands in Figures  \ref{2phBs}. 
As is well-known, the group velocity vector, $v_j^g$, and the energy-flux (Poynting) vector are coincident for the present class of problems, both pointing in the direction of energy flow.  
%%%%%%%%%
On each frequency band, $J$, the phase, $v^p_{Jj}$, and group, $v^g_{Jj} $, velocities are given by,
\begin{equation}\label{ph-group}
v^p_{Jj}=\frac{\omega_Jk_j}{k_1^2+k_2^2},\quad
v^g_{Jj}=\frac{\partial \omega_J}{\partial{k_j}},\quad
j=1,2;
\end{equation}
here, $J=1, 2,...$ denotes the frequency band and $j=1,2$.   
At each point on a frequency contour in Figures (\ref{2phBs}), one can readily identify the direction of the 
energy flow and the corresponding direction of the phase velocity. 

As is seen,  when the principal axes of the matrix dielectric tensor coincide with the $x_1, x_2$-axes, we see  on the first pass band
the group velocity vectors pointing outwards on the first quadrant, indicating positive refraction for both the phase velocity and the group velocity (energy flux). 

\begin{figure}%
\centering
\begin{minipage}[b]{0.45\linewidth}
\includegraphics[scale=0.45, trim=0cm 6.5cm 1cm 5cm, clip=true]{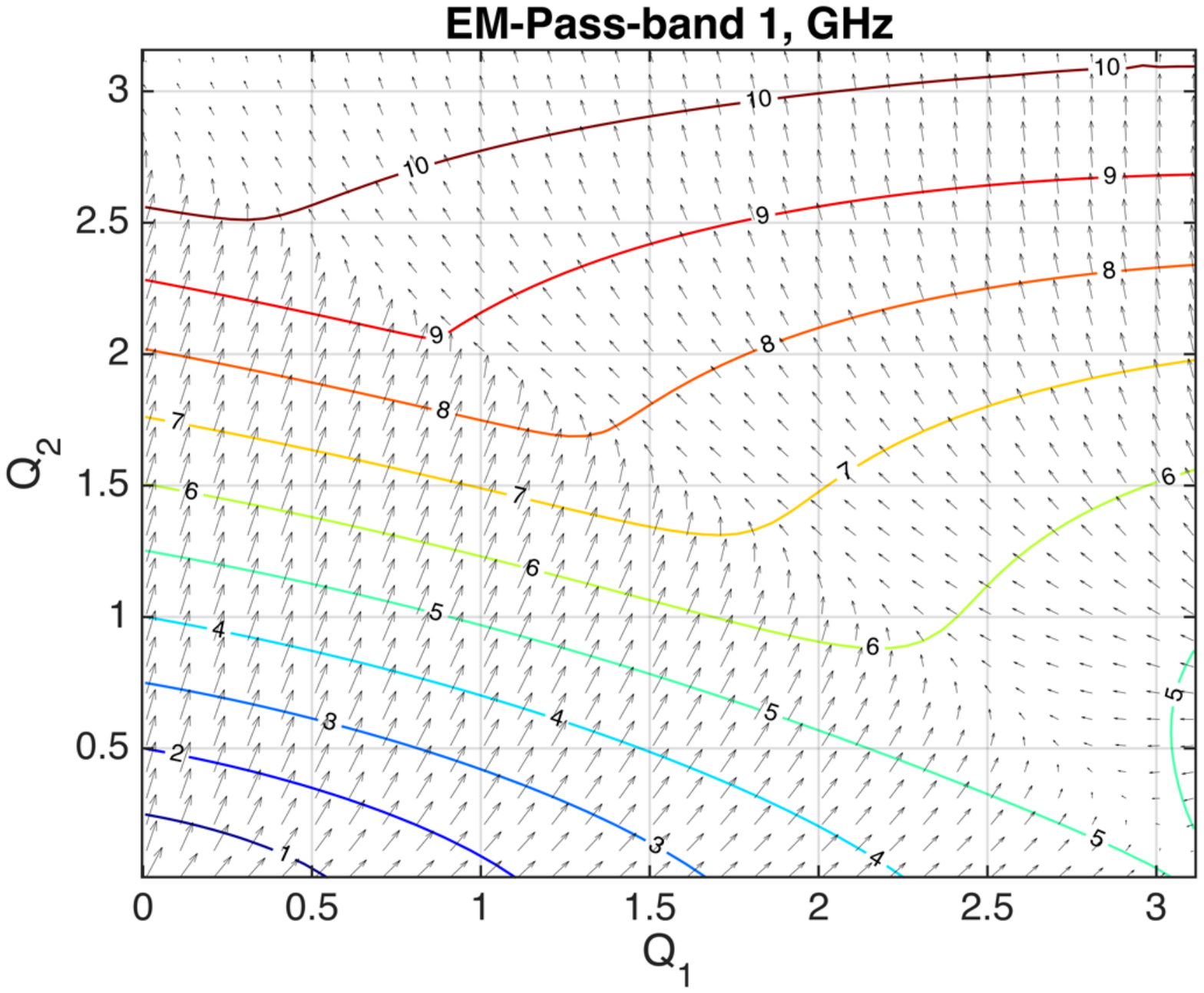}
\end{minipage}
\quad
\begin{minipage}[b]{0.45\linewidth}
\includegraphics[scale=0.45, trim=0.5cm 6.5cm 0cm 5cm, clip=true]{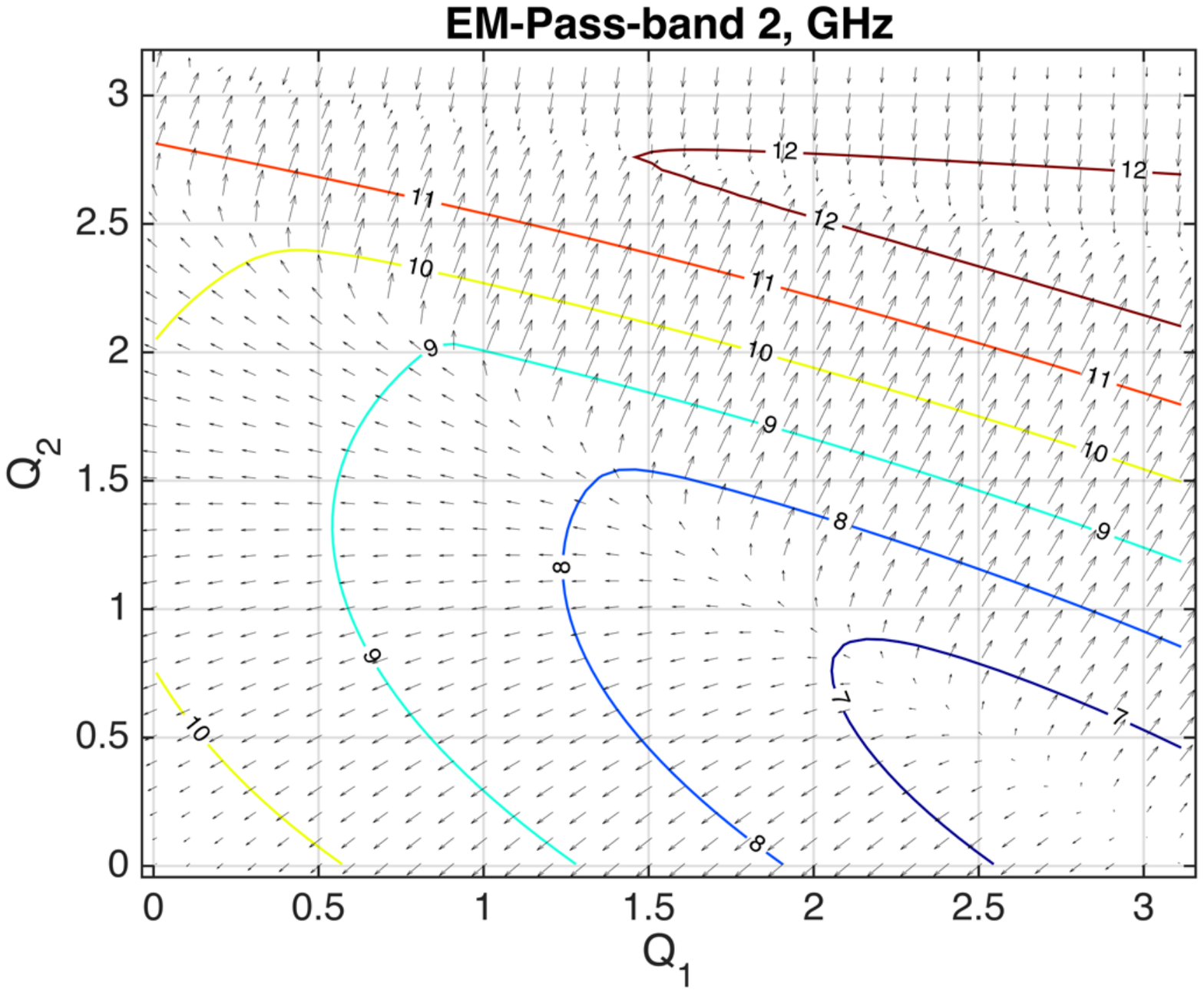}
\end{minipage}
\caption{Equi-frequency contours and energy-flux vectors of the first two pass bands of a photonic crystal with unit cell shown in Figure (\ref{unit_cells}b) for normalized wave vectors in the range  $0<Q_1,Q_2<\pi$; arrows are the energy-flux vectors.}
\label{FineC2phB}
\end{figure}

Here however, for the first time, without introducing a loaded series of capacitance and inductance \cite{siddiqui2003periodically}, we have realized negative energy refraction with positive phase-velocity refraction on a wide region of the first (lowest) frequency pass band of a photonic crystal of very simple unit cell: an anisotropic matrix with a central hole.
In the past, negative refraction has been reported on the second pass band of periodic composites, similar to what is seen on the second band in the left column graphs of Figures (\ref{2phBs}). Here, even the second pass band in the right column graphs of 
Figures (\ref{2phBs}) embodies a rich variety of refractive characteristics, including positive refraction over a broad range of frequencies. These new results are better seen in Figures (\ref{FineC2phB}) on the 
first (left) and second (right) pass bands of the unit cell in Figure (\ref{unit_cells}b)
for the normalized wave vector in the range $0<Q_1,Q_2<\pi$.
%

%%%%%%%%
\subsection{Three-phase Composite}
\begin{figure}%\label{3ph_10bands}%f5
\centering
\begin{minipage}[b]{0.45\linewidth}
\includegraphics[scale=0.4, trim=0cm 6.5cm 0cm 5cm, clip=true]{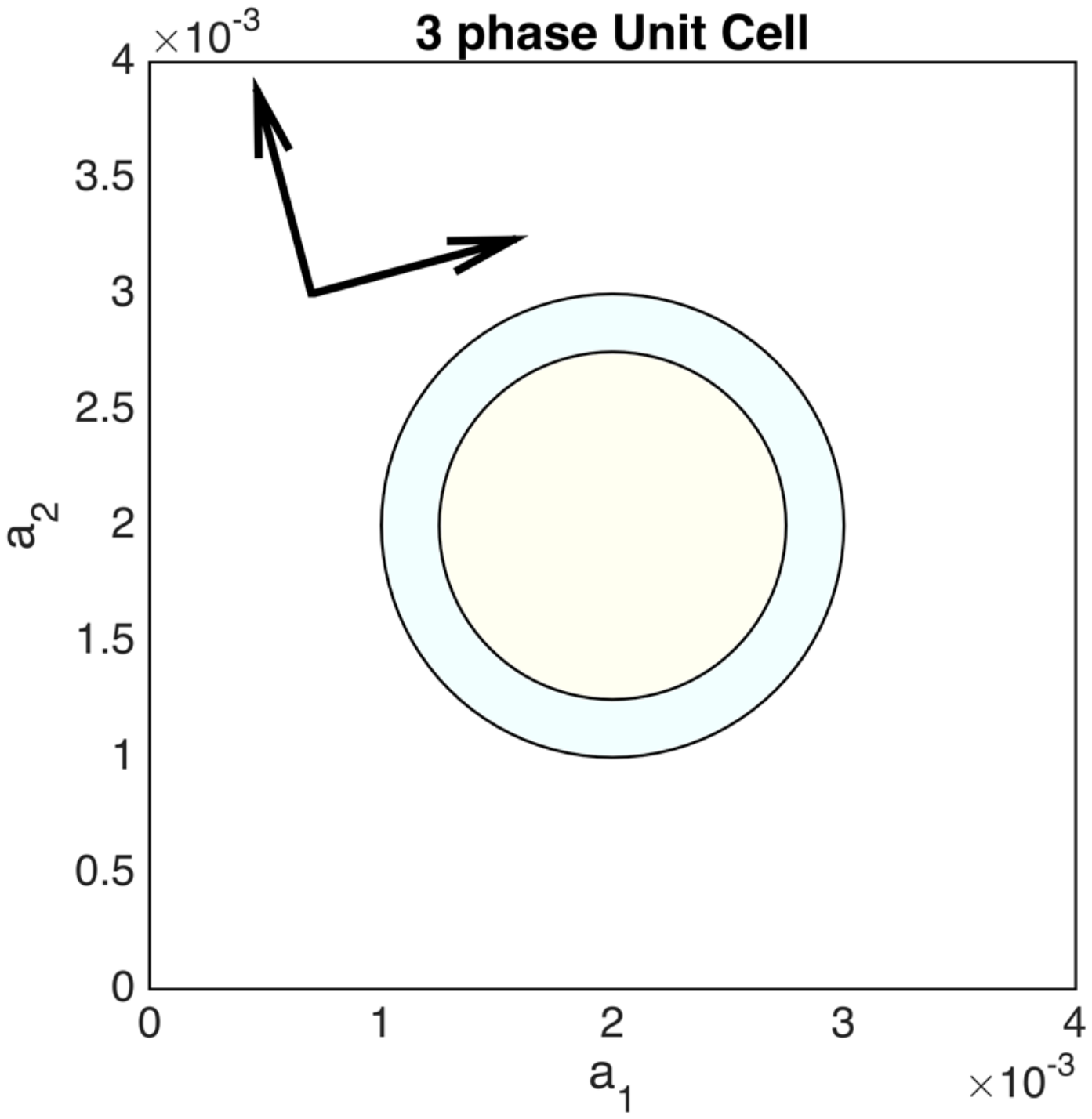}
\end{minipage}
\quad
\begin{minipage}[b]{0.45\linewidth}
\includegraphics[scale=0.4, trim=3cm 6.5cm 0cm 5cm, clip=true]{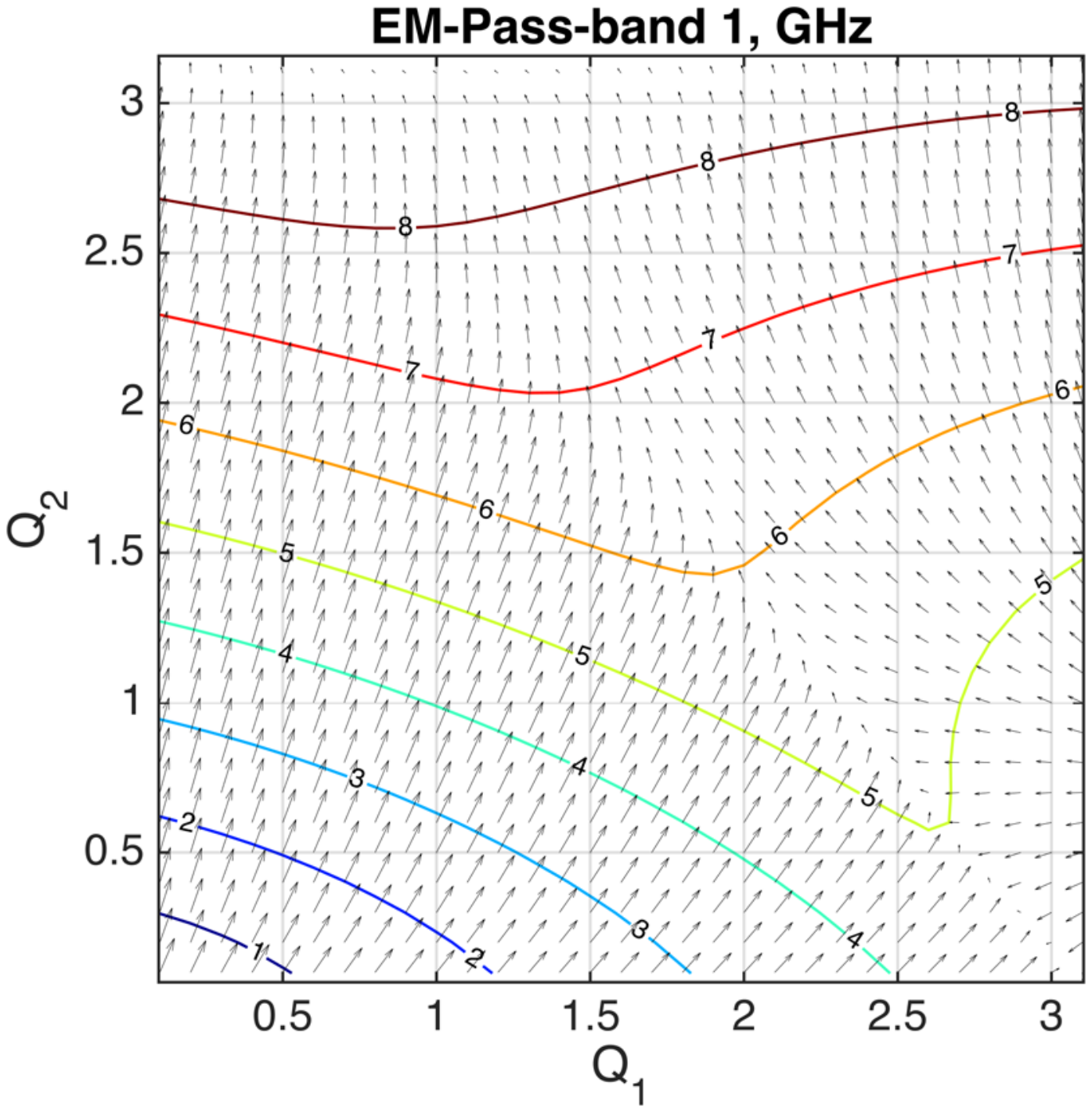}
\end{minipage}
\caption{(Left): Unit cell of a three-phase photonic crystal consisting of an anisotropic matrix with a 0.5mm high-dielectric material surrounding the central circular hole; (Right): the first pass band for $0< Q_1,Q_2<\pi$.  Principal axes of electric permittivity tensor of the matrix (Left, heavy arrows) are at about 15$^o$ angle with respect to the $x_1,x_2$-coordinate axes.}
\label{3ph}
\end{figure}

The computational tool presented in this paper allows to efficiently and accurately design unit cell for any desired refractive objectives. The available variables are the geometry of the unit cell, its composition, and in the present case the orientation of the principal axes of the matrix relative to the geometric symmetry of the unit cell.  To illustrate this possibility, we modifying the above-considered unit cell by encasing the central circular hole with a 0.5mm material of high dielectric constant, say, $\epsilon_{11}=\epsilon_{22}=150$, keeping all other parameters the same.  Dielectric solids with very high dielectric constant have been fabricated; see, e.g., \cite{sengupta1999breakthrough}, \cite{si2002epitaxial}.
The corresponding unit cell and a portion of the first pass band are shown in
Figures (\ref{3ph}) for $0<Q_1,Q_2<\pi$.  As is seen the frequencies are reduced by about  20\%. 
% &&&&&&&&&&&&&&&&&&&&&&&&&&&&&&&&&&&&&&&&&&&&&&&&&&&&

\subsection{Unit-cell Properties for Phononic Crystal; SH-waves}\label{phononic_prop}
\begin{figure}
\centering
\begin{minipage}{0.45\linewidth}
\includegraphics[scale=0.40, trim=0cm 1cm 0cm 0cm, clip=true]{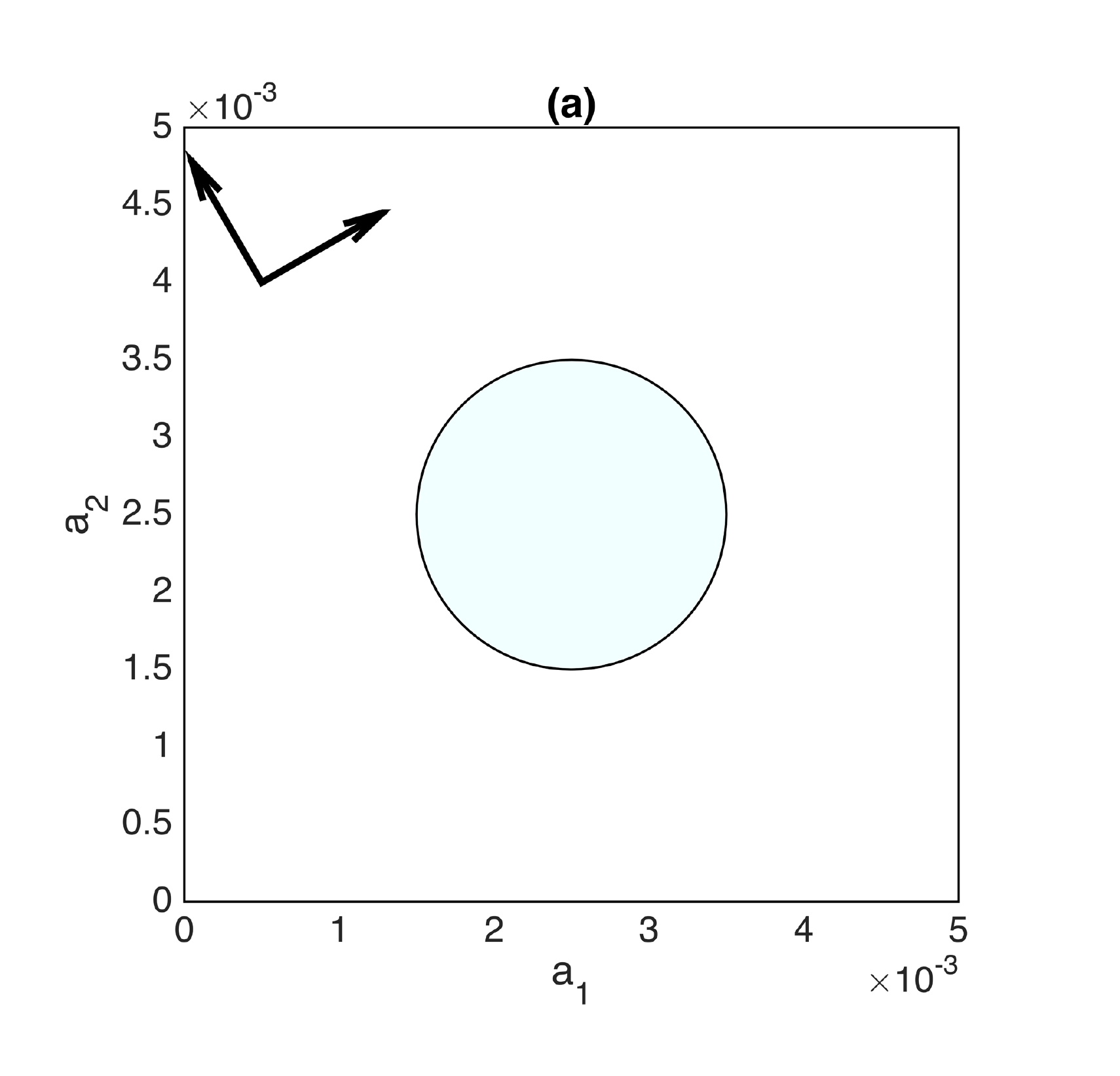}
\end{minipage}
%\quad
\begin{minipage}{0.45\linewidth}
\includegraphics[scale=0.40, trim=0cm 1cm 0cm 0cm, clip=true]{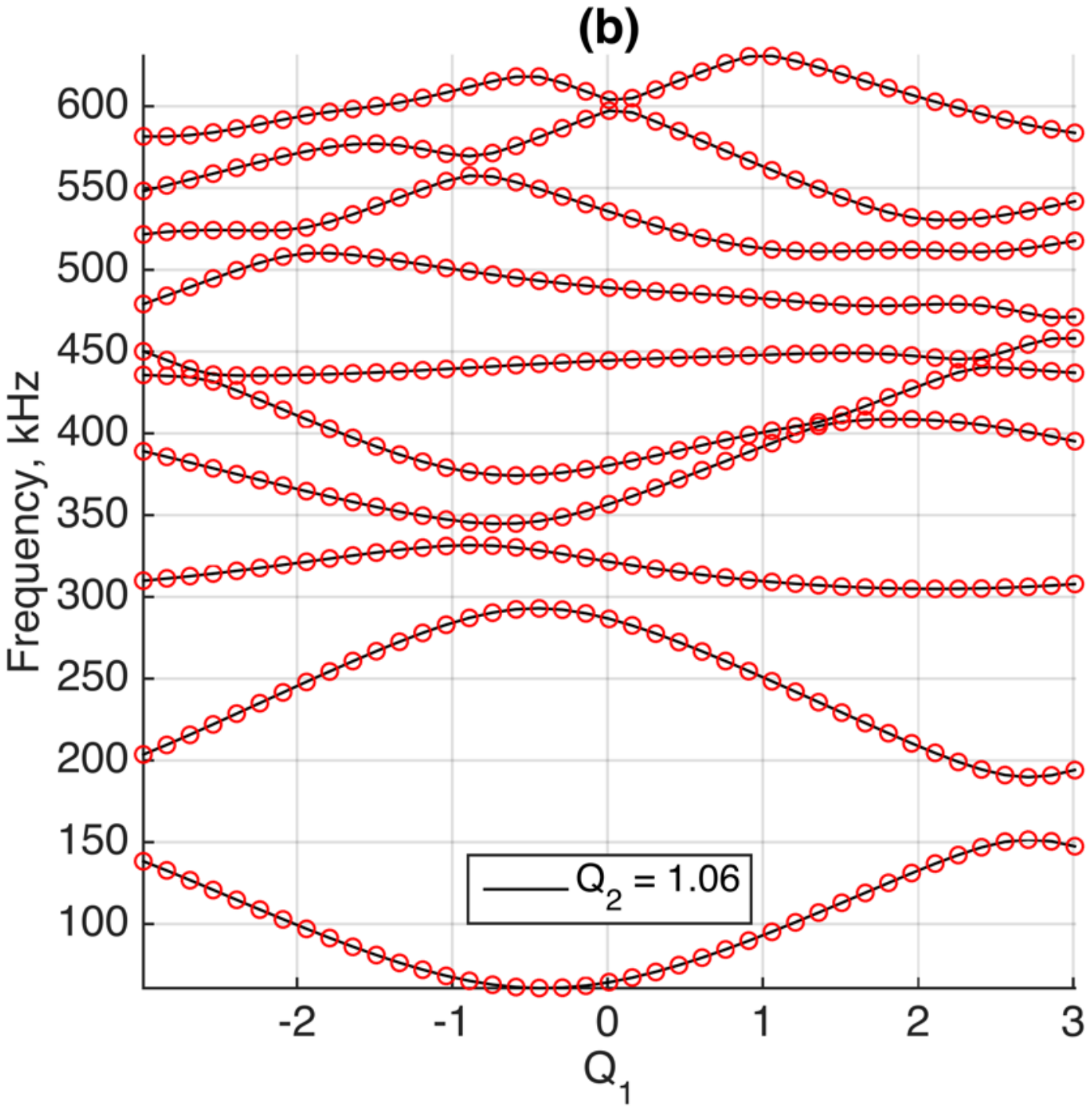}
\end{minipage}
\caption{(a): A two-phase phononic crystal.  (b): The first ten pass bands for $Q_2 = 1.06$; solid lines are by direct calculation using equations (\ref{SHB3}); open circles are by homogenized parameters using equation (\ref{HSH3}).}\label{FigSH}
\end{figure}
We consider a simple two-phase phononic crystal with the unit cell shown in Figure  (\ref{FigSH}a).  It consists of a unidirectionally reinforced carbon fiber-epoxy composite matrix containing 2mm-diameter circular aluminum-inclusions that are spaced  5mm center-to-center in the $x_1$ and $ x_2$-directions such that the principal directions of the matrix stiffness tensor make a 30$^o$ angle with the $x_1,x_2$-axis, as shown by the arrows in Figure (\ref{FigSH}a). The principal values of the shear modulus tensor are measured at 500kHz to be $\mu_I^{M} = 2.4GPa$ and $\mu_{II}^{M} = 5.8GPa$. The corresponding components in the  $x_1, x_2$-coordinates, and the properties of the aluminum inclusions are,

\begin{enumerate}
\item $\mu_{11}^{M}$=3.25GPa; $\mu_{22}^{M}$=4.95GPa; $\mu_{12}^{M}$=1.47GPa; $\rho^{M}$=1440kg/m$^3$.
\item $\mu_{11}^{In}=\mu_{22}^{In}$=28GPa; $\rho^{In}$=2700kg/m$^3$.
\end{enumerate}

\begin{figure}
\centering
\begin{minipage}{0.45\linewidth}
\includegraphics[scale=0.40, trim=0cm 4cm 0cm 5cm, clip=true]{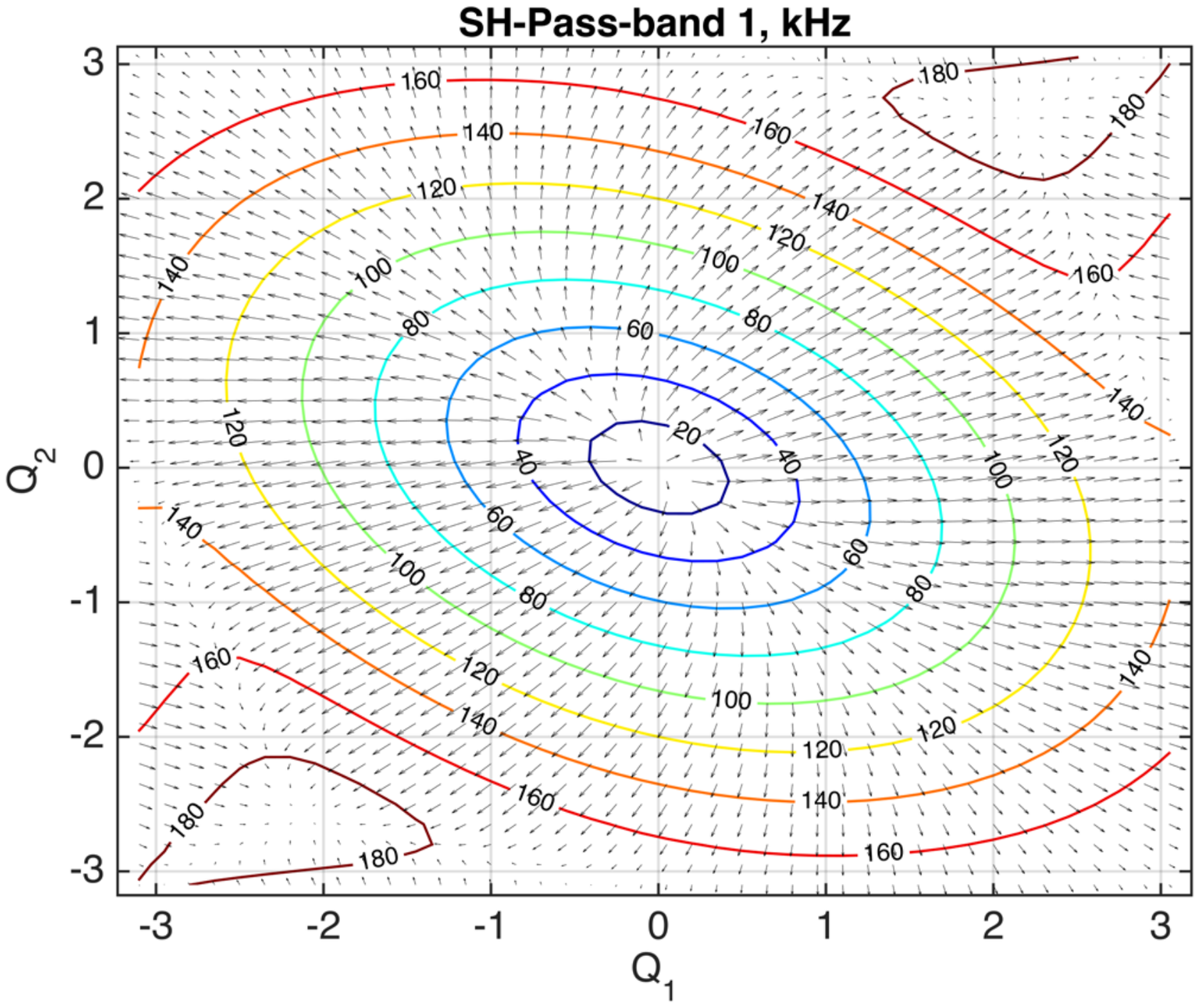}
\end{minipage}
%\quad
\begin{minipage}{0.45\linewidth}
\includegraphics[scale=0.40, trim=0cm 4cm 0cm 5cm, clip=true]{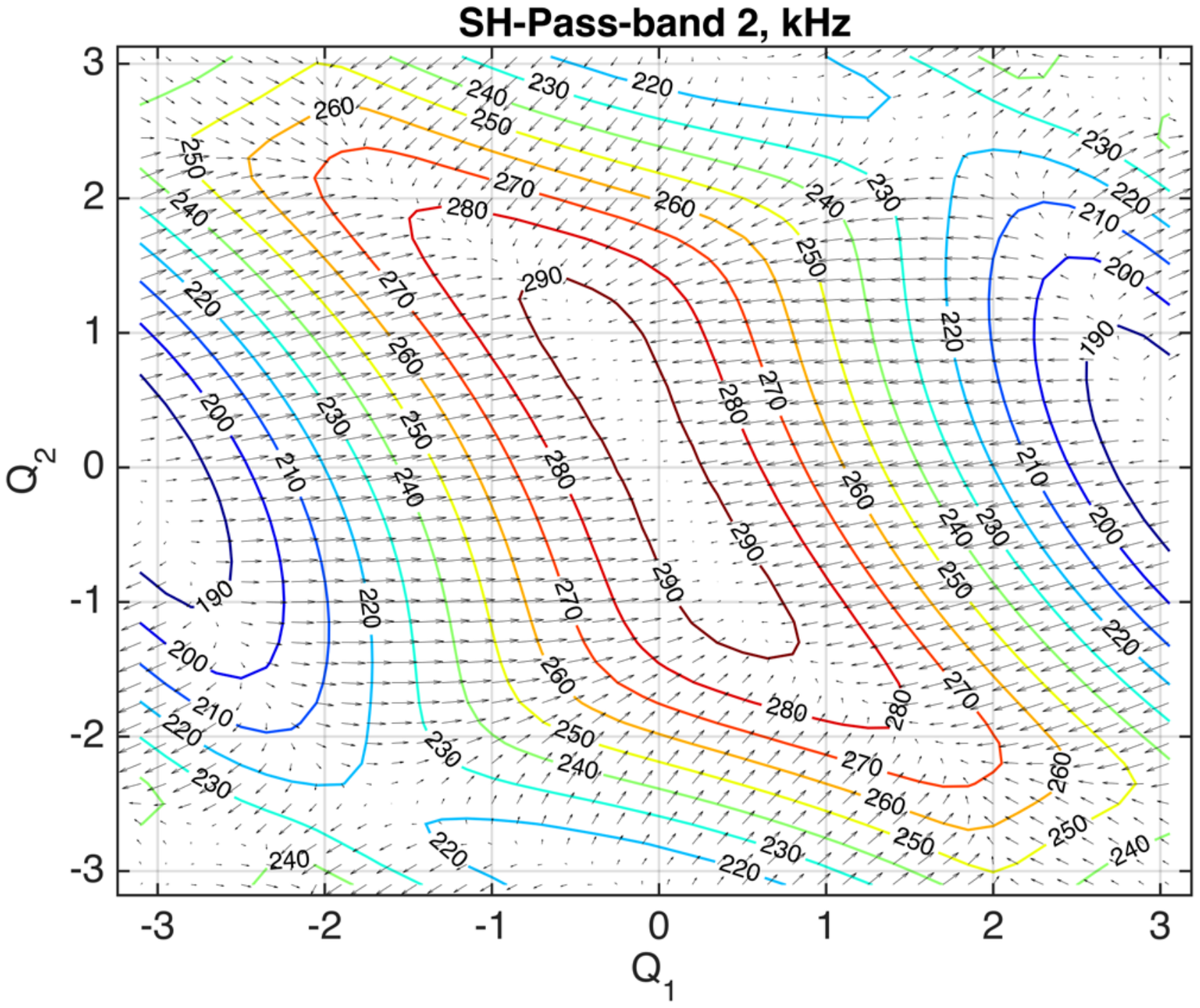}
\end{minipage}
%\quad
\begin{minipage}{0.45\linewidth}
\includegraphics[scale=0.40, trim=0cm 4cm 0cm 7cm, clip=true]{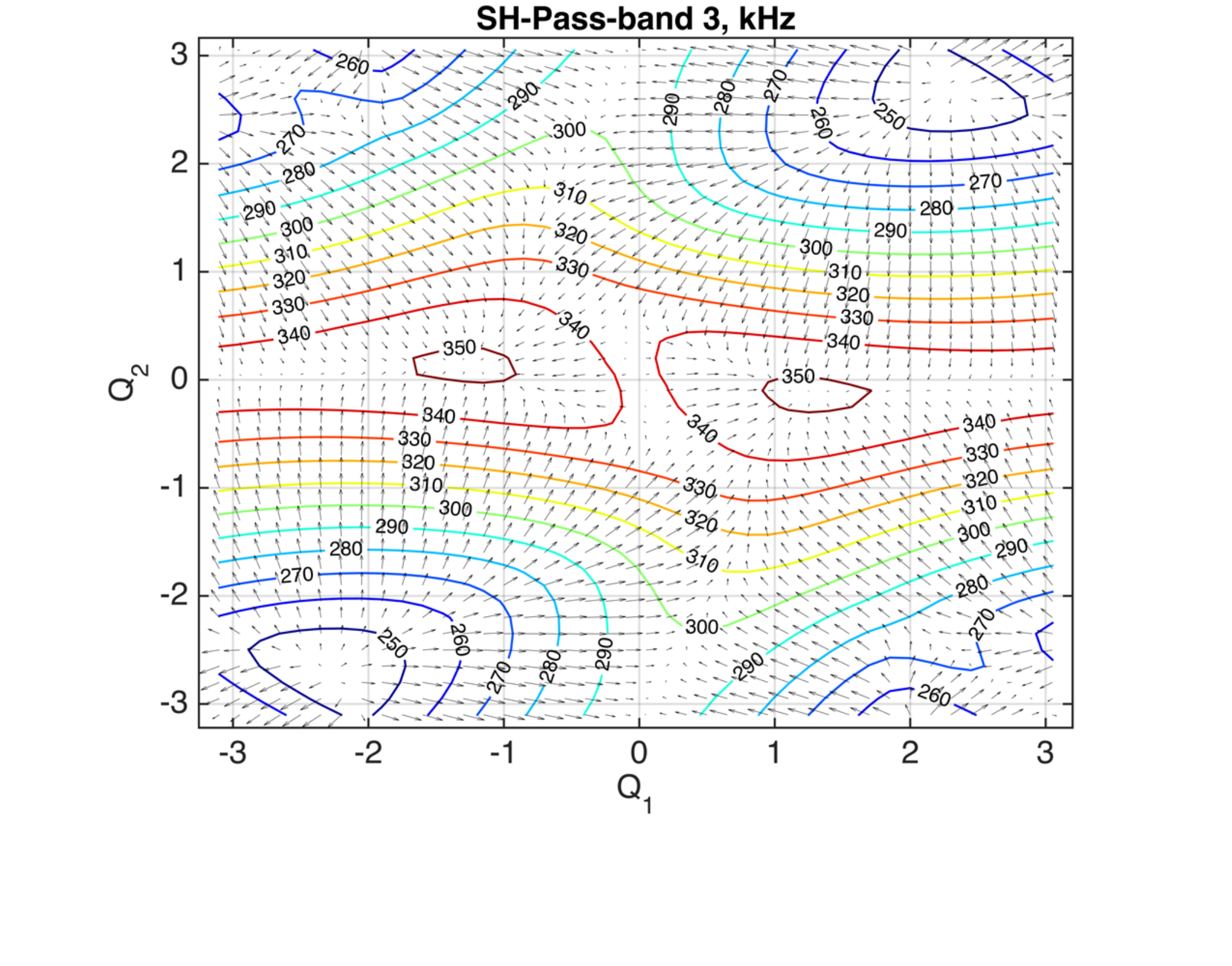}
\end{minipage}
\begin{minipage}{0.45\linewidth}
\includegraphics[scale=0.40, trim=0cm 4cm 0cm 7cm, clip=true]{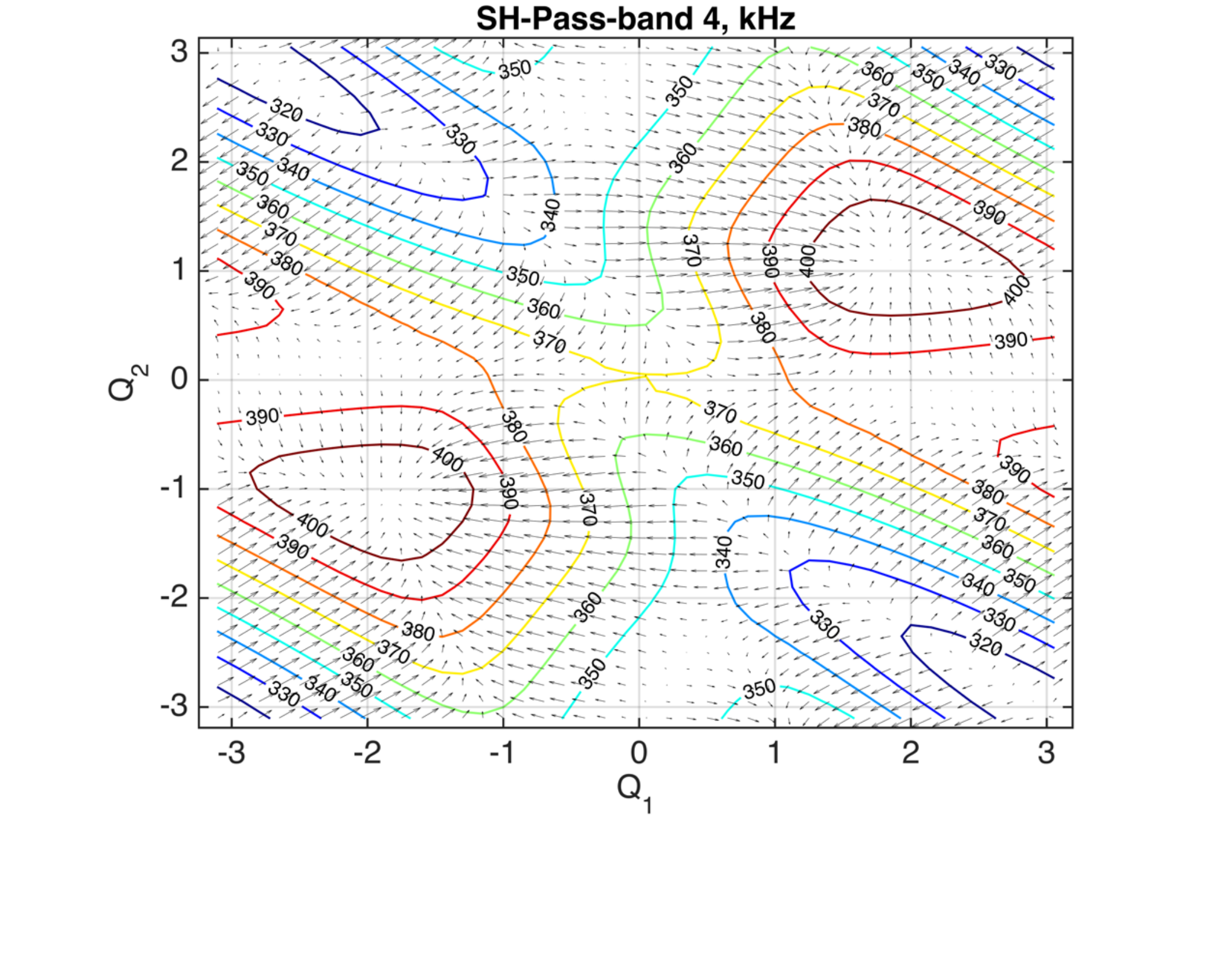}
\end{minipage}
\caption{Equi-frequency contours of the first four pass bands of the phononic crystal  
Figure (\ref{FigSH}a); arrows are the energy-flux vectors.}
\label{4phSH}
\end{figure}
The first ten pass bands of this crystal are shown in Figure  (\ref{FigSH}b) for $Q_2=1.06$, and in Figure
(\ref{4phSH}) we have shown the equi-frequency contours of the first four pass bands.
The arrows in the contour maps are  the corresponding energy-flux vectors.  

\begin{figure}%
\centering
\begin{minipage}{0.45\linewidth}
\includegraphics[scale=0.40, trim=0cm 4cm 0cm 5cm, clip=true]{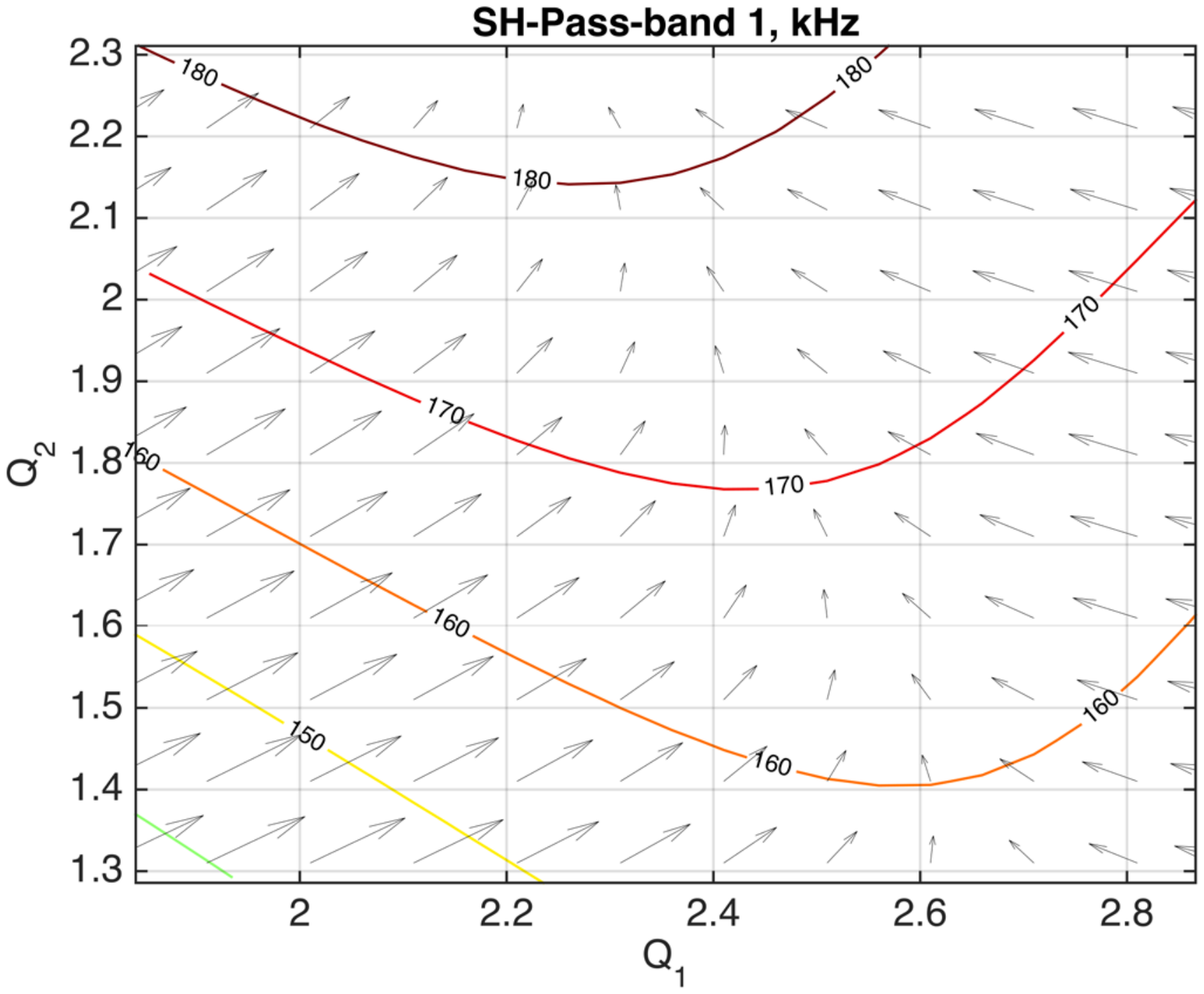}
\end{minipage}
%\quad
\begin{minipage}{0.45\linewidth}
\includegraphics[scale=0.40, trim=0cm 4cm 0cm 5cm, clip=true]{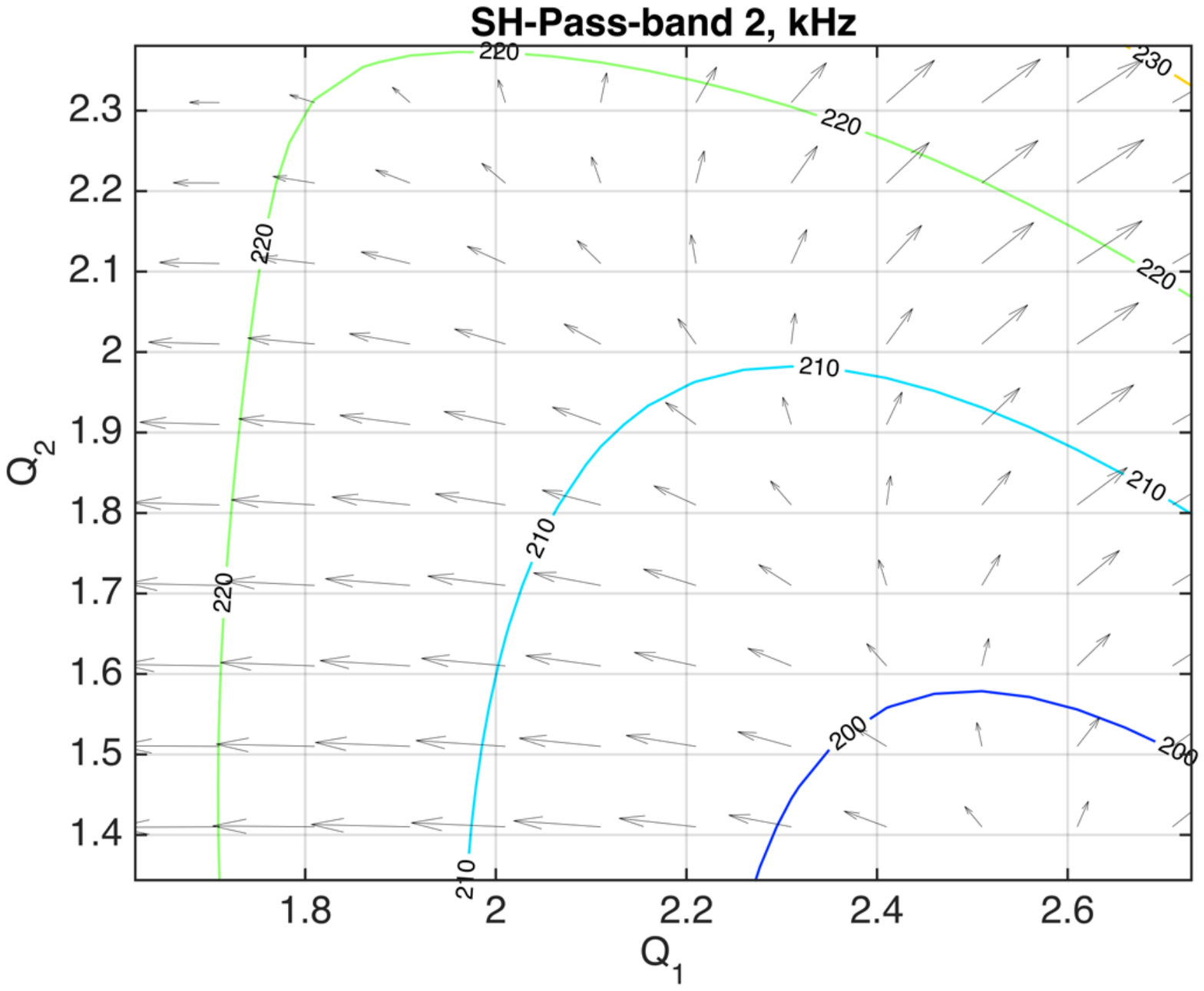}
\end{minipage}
\caption{Equi-frequency contours of phononic crystal with unit cell shown in Figure (\ref{FigSH}a) for indicated values of the wave-vector components over the first two (lowest two) pass bands where the phase velocity is positive; arrows are the energy-flux vectors.}
\label{FContSH2}
\end{figure}

As is evident from the energy-flux directions over the first (lowest) frequency pass band in  Figure (\ref{4phSH}), negative refraction occurs with positive phase-velocity refraction over a wide range of positive phase angles.  Indeed by simply suitably arranging the axes of the matrix anisotropy relative to the $x_1,x_2$-axes, a remarkably rich body of refraction characteristics of the composite is realized.   We believe these results and our proposed method of analysis are new, and should provide powerful tools for designing novel phononic crystals.
In Figures (\ref{FContSH2}) we have captured the refraction response of this simple phononic crystal for the indicated values of the wave-vector components over the first two (lowest two) pass bands where the phase velocity is positive. Positive or negative energy refraction is captured at various frequencies.
%$$$$$$$$$$$$$$$$$$$$$$$$$$$$$$$$$$$$$$$$$$$

\section{Discussion and Conclusions}\label{conclusions}

We have shown, for the first time, that negative refraction with positive phase velocity refraction can be realized over a broad range of frequencies on the first (lowest) frequency pass band of very simple photonic and phononic crystals, and that this can be controlled by suitable unit-cell design.  

First we have provided a unified methodology to accurately and efficiently produce the  homogenized effective material properties of doubly periodic phononic crystals for anti-plane shear (SH) and photonic crystals for transverse electric (TE) and transverse magnetic (TM) electromagnetic Bloch-form waves over any desired  range of frequencies.
We have shown the correspondence between phononic and photonic field equations, and have outlined in detail the actual calculation procedures. We have worked out illustrative examples for each case and have unveiled a rich body of refractive characteristics of these crystals that have thus far been hidden. 
%%%%%%%%%%%%%%%%%%%

It turns out that both photonic and phononic crystals with even the simplest unit cells have a rich body of refractive responses that can be realized even on the first (lowest) frequency pass band. Indeed, on the first and second frequency pass bands, one can have:
(a) negative refraction with negative phase-velocity refraction,
(b) negative refraction with positive phase-velocity refraction,
(c) positive refraction with negative phase-velocity refraction,
(d) positive refraction with positive phase-velocity refraction, or even
(e) complete reflection with no energy transmission,
depending on the frequency and wave vector.

Our calculation method applies to any
unit cell that may consist of anisotropic constituents of any constant or variable properties that may also admit large discontinuities.
The homogenized (spatially constant) effective material parameters depend parametrically on the wave frequency and wave vector.   \textit{They embody the actual dispersive properties of the considered phononic/photonic crystal, and produce their exact band structure.}
Using the same volume fraction of the same constituents, one can design a (homogenized) material that would display negative energy refraction with positive phase velocity refraction even on its first (acoustic) pass band, providing opportunity for a variety of applications where devices with low loss refractive properties are required.
The window of  frequency and the range of wave-vector values at which the negative energy refraction would occur can be controlled by judicious selection and design of the unit cell's constituent geometry and properties.

\textbf{Acknowledgments}:
This research has been conducted at the Center of Excellence for Advanced Materials (CEAM) at the University of California, San Diego, under DARPA  RDECOM W91CRB-10-1-0006 to the University of California, San Diego.

\appendix

\section{Band Structure Calculation}\label{AppendixA}

The calculation of the band structure and the associated field variables for two- and three-dimensional phononic/photonic crystals presents a formidable challenge especially when their unit cells 
consist of constituents of diverse properties with large discontinuities. There are limited computational tools that can be employed, each with its own limitations, see for example the following and cited references therein:
\cite{leung1990full},
\cite{plihal1991photonic},
\cite{kushwaha1993acoustic},
\cite{pendry1994photonic}, 
 \cite{bell1995program}, 
 \cite{tran1995photonic},
\cite{pendry1996calculating},
\cite{chou2014eigendecompositions},
 \cite{han2014numerical}.
In the present work, we use a mixed variational formulation for band-structure calculations, where both the displacement and stress fields in the phononic and the electric and magnetic fields in the photonic crystals are varied independently.  Hence they may be approximated by any  continuously differentiable  set of complete base functions, even though the gradients of some of the field variables may suffer large discontinuities across interfaces of various constituents of a typical unit cell. 
Since the method  is based on a variational principle, any set of approximating functions can be used for calculations, e.g., plane-wave Fourier series or finite elements \cite{nemat1975harmonic}, 
\cite{minagawa1976harmonic}, \cite{minagawa1981finite}. 
The mixed variational produces very accurate results and the rate of convergence of the corresponding series solution is superior to alternative methods \cite{babuska1978}. The general approach has been used to develop a fast and accurate computational platform for band structure calculations \cite{lu2016variational}, where it is also shown that the mixed variational approach provides greater convergence rate than the usual Rayleigh quotient, especially in the presence of large material discontinuities.  
Other variational methods have also been used for band structure calculation of periodic composites, see, e.g., 
\cite{goffaux2003two}. 

In what follows,  detailed calculations are given for the SH waves and then the results for TE and TM waves are extracted using the corresponding equations.

\subsection{SH Waves} 

We minimize the following functional, considering the displacement and stresses as independent fields subject to variation:
\begin{equation}\label{SHMIX}%22
\omega^2=
\frac{<\tau_j,{w}_{,j}>+<{w}_{,j},\tau_j>-<D_{jk} \tau_k,\tau_j>}{ <\rho {w},{w}>};\quad
j,k=1,2,
\end{equation}
where $<g u,v> = \int_{-1/2}^{1/2}\int_{-1/2}^{1/2}  guv^*dx_1dx_2$ for a real-valued function $g(x_1,x_2)$ and complex-valued functions $u(x_1,x_2)$ and $v(x_1,x_2)$, with star denoting complex conjugate, the repeated indexes being summed. 

The functional (\ref{SHMIX}) was introduced by the present author in the early 1970's (\citeasnoun{nemat1972harmonic}) and was termed \textit{new quotient}.  It yields remarkably accurate frequency bands  of phononic crystals using only a few plane-wave approximations as compared with the usual Rayleigh quotient \cite{nemat2015refraction}.
Here we use the following approximating functions:
\begin{eqnarray}\label{SH3}%23
\left[ \begin{array}{c}
w\\
\tau_j\\
\end{array} \right]=\sum_{n_1,n_2=-N}^{+N}
\left[ \begin{array}{c}
W^{n_1 n_2}\\
T_j^{n_1 n_2}\\
\end{array} \right]e^{i[(k_1+\frac{2\pi n_1}{a_1})x_1+(k_2+\frac{2\pi n_2}{a_2})x_2]},
\end{eqnarray}
which automatically ensures the Bloch and continuity conditions,
and, by direct substitution into (\ref{SHMIX}) and minimization with respect to $W$ and $T_j$ as the independent variables, obtain
\begin{eqnarray}\label{SHB1}%24
\left[ \begin{array}{ccc}
iH_1 & iH_2 & \omega^2 \Lambda_{\rho} \\
\Lambda_{D_{11}} & \Lambda_{D_{12}}& -iH_1\\
\Lambda_{D_{21}} & \Lambda_{D_{22}} & -iH_2
\end{array} \right]\
\left[ \begin{array}{c}
T_1\\
T_2\\
W\end{array} \right]=0,
\end{eqnarray}
where $\Lambda_f=[\Lambda^{(n_1 n_2,m_1 m_2)}_{f}]$ is an 
$M^2 \times M^2$ matrix, and
$H_{1}$ and $H_{2}$ are two $M^2\times M^2$ diagonal matrices with the respective components 
$(k_1+2\pi n_1)\delta_{n_1 m_1}\delta_{n_2 m_2}$ and
$(k_2+2\pi n_2)\delta_{n_1 m_1}\delta_{n_2 m_2}$,
where $M=2N+1$.
The components of $\Lambda_f$ are defined by
\begin{equation}\label{SHB2}%25
\Lambda^{(n_1 n_2,m_1 m_2)}_{f}=
\int_{-a_1/2}^{a_1/2}\int_{-a_2/2}^{a_2/2} f(x_1,x_2) 
e^{i2\pi [(n_1 - m_1)x_1+(n_2-m_2)x_2]}dx_1 dx_2,
\end{equation}
with $f(x_1,x_2)$ being a real-valued integrable function.  
For an even function, $f(x_1,x_2)=f(-x_1,-x_2)$ (symmetric unit cells), 
and 
$\Lambda^{(n_1 n_2,m_1 m_2)}_{f}=
\Lambda^{(m_1 m_2,n_1 n_2)}_{f}$ is real-valued.

From the system of linear and homogeneous equations (\ref{SHB1}),we obtain,
\begin{eqnarray}\label{SHB3}%26
\begin{array}{c}
\left[
\Phi-\omega^2\Omega
\right]W=0,\quad
\Omega=\Lambda_{\rho},\quad
\Phi=
H_j M_{jk} H_k,\quad
\left[M_{jk} \right]=
\left[\Lambda_{D_{jk}} \right]^{-1},
\end{array}
\end{eqnarray}
where $j,k=1,2$, and repeated indices are summed.
For given values of $k_1$ and $k_2$, the eigenvalues, $\omega$, of equation(\ref{SHB3})$_1$ are obtained from 
\begin{equation}\label{SHB4}%27
det\left|\Phi
-\omega^2\Omega\right|=0,
\end{equation}
and for each eigenvalue, the corresponding $M\times M$ displacement matrix, $W=[W^{n_1 n_2}]$, is given by (\ref{SHB3})$_1$,  and the stresses, $T_j=[T^{n_1 n_2}_j]$, and strains, $\gamma_j=[\gamma^{n_1 n_2}_j]$,
 by,
\begin{equation}\label{SHB5}%28 
T_{j}=iM_{jk}H_k W,\quad
\gamma_{j}=\Lambda_{D_{jk}}T_k,\quad
\gamma_{j}= iH_jw.
\end{equation}

%%%%%%%%%%%%%%%%%%%%%%%%%%%%%%%%%%%%%%%%%%%

\subsection{TE/TM  Waves} 

For the TE Bloch-form waves, set
\begin{eqnarray}\label{HTE29}%29
\left[ \begin{array}{c}
\mathcal{H}\\
\mathcal{E}_j\\
\end{array} \right]=\sum_{n_1,n_2=-N}^{+N}
\left[ \begin{array}{c}
\mathcal{H}^{n_1 n_2}\\
\mathcal{E}_j^{n_1 n_2}\\
\end{array} \right]e^{i[(k_1+\frac{2\pi n_1}{a_1})x_1+(k_2+\frac{2\pi n_2}{a_2})x_2]},
\end{eqnarray}
and for the TM Bloch-form waves, set 
\begin{eqnarray}\label{HTM11}%30
\left[ \begin{array}{c}
\mathcal{E}\\
\mathcal{H}_j\\
\end{array} \right]=\sum_{n_1,n_2=-N}^{+N}
\left[ \begin{array}{c}
\mathcal{E}^{n_1 n_2}\\
\mathcal{H}_j^{n_1 n_2}\\
\end{array} \right]e^{i[(k_1+\frac{2\pi n_1}{a_1})x_1+(k_2+\frac{2\pi n_2}{a_2})x_2]}.
\end{eqnarray}
Then using the correspondence relations (\ref{SHTE}), we obtain, for TE-waves,
\begin{eqnarray} \label{TEB1}%31
\begin{array}{lr}
[\Phi-\omega^2\Omega]\mathcal{H}=0,~~
\Omega=\Lambda_{\mu},\\
\Phi=H_2\nu_{11}H_2+H_1\nu_{22}H_1-
(H_2\nu_{12}H_1+
H_1\nu_{21}H_2),~~\\
{\mathcal{E}}_1=\frac{1}{\omega}
(-\nu_{11}H_2+\nu_{12}H_1)\mathcal{H},~~~\\
{\mathcal{E}}_2=\frac{1}{\omega}
(-\nu_{21}H_2+\nu_{22}H_1)\mathcal{H};~~~
[\nu_{jk}]=[\Lambda{\epsilon_{jk}}]^{-1}.
\end{array}
\end{eqnarray}
Similarly, for TM-waves, we use (\ref{SHTM}) and obtain,
\begin{eqnarray}\label{TMB1}%32
\begin{array}{lr}
[\Phi-\omega^2\Omega]\mathcal{E}=0,~~~
\Omega=\Lambda_{\epsilon},\\
\Phi=H_2\lambda_{11}H_2+H_1\lambda_{22}H_1-
(H_1\lambda_{21}H_2+H_2\lambda_{12}H_1)\mathcal{H},\\
{\mathcal{H}}_1=\frac{1}{\omega}
(\lambda_{11}H_2-\lambda_{12}H_1)\mathcal{E},\\
{\mathcal{H}}_2=\frac{1}{\omega}
(\lambda_{21}H_2-\lambda_{22}H_1)
\mathcal{E},~~~
[\lambda_{jk}]=[\Lambda{\mu_{jk}}]^{-1},
\end{array}
\end{eqnarray}
where $\Lambda{\epsilon_{jk}}$ and $\Lambda{\mu_{jk}}$ are defined by equation (\ref{SHB2}); see also below.
%%%%%%%%%%%%%%%%%%%%%%%%%%

\subsection{Expressions for $\Lambda_{f(x_1,x_2)}$  in Special Cases}

When a rectangular unit cell contains a nested sequence of either rectangular or elliptical inclusions, matrix $\Lambda_{f(x_1,x_2)}$ can be  calculated explicitly for piecewise constant values of $f(x_1,x_2)$.  Consider an $a_1$ by $a_2$ unit cell, $\Omega_1$, that contains a nested sequence of $n-1$ concentric either elliptical or rectangular subregions, 
$ \Omega_1\supset \Omega_2 \supset \Omega_3 ~  {...}  \supset \Omega_n.$

Denote the dimensions of the principal axes of a typical subregion $\Omega_j$ by ${a}_1(j)$ and ${a}_2(j)$. 
Then the area of the $j^{th}$ subregion is $\Omega_j-\Omega_{j-1}$, where for
a rectangular subregion 
 $\Omega_j=a_1(j)a_2(j)$, and for an elliptical subregion 
$\Omega_j=\frac{\pi}{4}a_1(j)a_2(j)$.
Let $f(j)$ stand for either the mass-density or the shear modulus of the subregion $\Omega_j-\Omega_{j-1}$.  From (\ref{SHB2}) now obtain,
\begin{equation}\label{Lambda}%33
   \Lambda_f=
	 \sum_{k=2}^{n}(f_k-f_{k-1})g_k,
\end{equation}
\begin{equation}\label{g_k1} %34
	g_k=\int_{\Omega_k}
	 exp\{i2\pi[(\frac{n_1-m_1}{a_1})x_1+(\frac{n_2-m_2}{a_2})x_2]\} dx_1dx_2, 
\end{equation}
where for a rectangular subregion $g_k$ is given by,
\begin{equation}%35
 g_k =\begin{cases}
 	 \frac{sin(\pi nm_1a_1(k)/a_1)}{\pi nm_1} 
	  \frac{sin(\pi nm_2 a_2(k)/a_2)}{\pi nm_2} & nm_1\ne 0,\quad nm_2\ne 0,\\
        \frac{sin(\pi nm_1a_1(k)/a_1)}{\pi nm_1 a_2}a_2(k) &  nm_1\ne 0,\quad nm_2 = 0,\\
        \frac{sin(\pi nm_2 a_2(k))}{\pi nm_2a_1}a_1(k)&  nm_1= 0,\quad nm_2 \ne 0,\\ 
        \frac{a_1(k)a_2(k)}{a_1a_2}, &  nm_1= 0,\quad nm_2 = 0,\\  
nm_1=n_1-m_1,~ nm_2=n_2-m_2,\\
(k ~ not ~ summed);
 \end{cases}
\end{equation}
and for an elliptical subregion $g_k$ becomes,
\begin{equation}%36
g_k=\frac{\pi}{2}\frac{a_1(k)a_2(k)J_1(R_k)}{a_1a_2R_k},\quad
R_k=\pi \{[nm_1a_1(k)/a_1]^2+[nm_2a_2(k)/a_2]^2\}^{1/2}.
\end{equation}

%$$$$$$$$$$$$$$$$$$$$$$$$$$$$$$$$$$$$$$$
%$$$$$$$$$$$$$$$$$$$$$$$$$$$$$$$$$$$$$$$$$$$

\subsection{Calculation of Effective Properties}

The periodic part of the field variables is directly averaged to obtain the corresponding effective parameters according to equations (\ref{direct}) for SH waves and 
equations (\ref{EFFECTIVE3}) and (\ref{EFFECTIVE4}) for TE and TM. 
Here we give the equations for the calculation of the effective properties for SH-waves as illustration; similar equations hold for TE and TM, EM-waves,  
\begin{eqnarray} \label{HSH5}%37
 \begin{aligned}
&\overline{(D_{jk}\tau_k)}=D_{jk}^{eff}\bar{\tau}_{k}=\sum_{n_1,n_2=-N}^{+N} 
K^{n_1 n_2}[ D_{jk}(\mathbf{x})]T^{n_1 n_2}_k,\quad
\bar{\tau}_j=T^{0 0}_j,\\
&\overline{(\rho w)}=
\sum_{n_1,n_2=-N}^{+N} K^{n_1 n_2}[ \rho(\mathbf{x})]W^{n_1 n_2},\quad
\bar{w}=W^{0 0},\\
\end{aligned}
\end{eqnarray} 
where
\begin{equation}\label{HSH6}%38
K^{n_1 n_2}[f(\mathbf{x})]=K_{f(x_1,x_2)}=
\int_{-a_1/2}^{a_1/2}\int_{-a_2/2}^{a_2/2}f(\mathbf{x})
e^{i2\pi(\frac{n_1x_1}{a_1}+\frac{n_2x_2}{a_2}) }dx_1dx_2.
\end{equation}

As for $\Lambda_{f(x_1,x_2)}$, explicit expression can be obtained for  $K_{f(x_1,x_2)}$ 
when a unit cell contains a nested sequence of elliptical or rectangular inclusion.  Using a similar notation as in equation (\ref{Lambda}), we set,

\begin{equation}\label{Kmatrix}%39
K_f=
	 \sum_{k=2}^{n}(f_k-f_{k-1})h_k,
\end{equation}%40
\begin{equation}\label{h_k1} 
	h_k=\int_{\Omega_k}
	 exp\{i2\pi[\frac{n_1x_1}{a_1}+\frac{n_2x_2}{a_2}]\} dx_1dx_2, 
\end{equation}
where for a rectangular subregion $h_k$ is given by,
\begin{equation}%41
 h_k =\begin{cases}
 % \begin{array}{lr}
 	 \frac{sin(\pi n_1a_1(k)/a_1)}{\pi n_1} 
	  \frac{sin(\pi n_2 a_2(k)/a_2)}{\pi n_2} & n_1\ne 0,\quad n_2\ne 0,\\
        \frac{sin(\pi n_1a_1(k)/a_1)}{\pi n_1 a_2}a_2(k) &  n_1\ne 0,\quad n_2 = 0,\\
        \frac{sin(\pi n_2 a_2(k))}{\pi n_2a_1}a_1(k)&  n_1= 0,\quad n_2 \ne 0,\\ 
        \frac{a_1(k)a_2(k)}{a_1a_2}, &  n_1= 0,\quad n_2 = 0,\\  
   \end{cases}
\end{equation}

and for an elliptical subregion $h_k$ becomes,
\begin{equation}%42
h_k=\frac{\pi}{2}\frac{a_1(k)a_2(k)J_1(R_k)}{a_1a_2R_k},\quad
R_k=\pi \{[n_1a_1(k)/a_1]^2+[n_2a_2(k)/a_2]^2\}^{1/2}.
\end{equation}

\section{Averaging Procedure}\label{AppendixB}

Here we present the details of the averaging procedure.  
Consider the dynamic field equations (\ref{SH1}) and, for notational simplicity, introduce the two-dimensional vectors, 
$\mathbf{k, x, X}$,and $\mathbf{y}$ with the respective components, 
$k_j, x_j, X_j$, and $y_j$.  
Now multiply (\ref{SH1})$_{1,2,3}$ by $e^{-i\mathbf{k.X}} $ and write the resulting equations as,
\begin{eqnarray}\label{A1}
\begin{aligned}
&\frac{\partial}{\partial{x_j}}\big{(}\tau^p_{j}(\mathbf{x})e^{i\mathbf{k.( x-X)}}\big)+
\omega^2 \rho(\mathbf{x}) w^p(\mathbf{x})e^{i\mathbf{k.(x-X )}}=0,\\
&\frac{\partial}{\partial{x_j}}\big{(}{w}^p(\mathbf{x})e^{i\mathbf{k. (x-X)}}\big)
-\gamma^p_j(\mathbf{x})e^{i\mathbf{k. (x-X)}}=0,\\
&\tau^p_j(\mathbf{x})e^{i\mathbf{k. (x-X)}}-
\mu_{jk}(\mathbf{x})\gamma^p_ke^{i\mathbf{k. (x-X)}}=0,\\
&\gamma^p_j(\mathbf{x})e^{i\mathbf{k. (x-X)}}-
D_{jk}(\mathbf{x})\tau^p_ke^{i\mathbf{k. (x-X)}}=0,\\
\end{aligned}
\end{eqnarray} 
set $ \mathbf{y}=\mathbf{x}-\mathbf{X}$ and obtain,
\begin{eqnarray} \label{A2}
 \begin{aligned}
&\frac{\partial}{\partial{y_j}}\big{(}\tau^p_{j}(\mathbf{X}+\mathbf{y})e^{i\mathbf{k. y}}\big)+
\omega^2\rho(\mathbf{X}+\mathbf{y}) \big{(}w^p(\mathbf{X}+\mathbf{y})e^{i\mathbf{k. y}}\big)=0,\\
&\frac{\partial}{\partial{y_j}}\big{(}{w}^p(\mathbf{X}+\mathbf{y})e^{i\mathbf{k. y}}\big)
- \gamma^p_j(\mathbf{X}+\mathbf{y})e^{i\mathbf{k. y}}=0,\\
&\tau^p_j(\mathbf{X}+\mathbf{y})e^{i\mathbf{k. y}}-
\mu_{jk}(\mathbf{X}+\mathbf{y})\gamma^p_k(\mathbf{X}+\mathbf{y})e^{i\mathbf{k. y}}=0,\\
&\gamma^p_j(\mathbf{X}+\mathbf{y})e^{i\mathbf{k. y}}-
D_{jk}(\mathbf{X}+\mathbf{y})\tau^p_k(\mathbf{X}+\mathbf{y})e^{i\mathbf{k. y}}=0.\\
\end{aligned}
\end{eqnarray} 
Finally average the field equations with respect to$\mathbf{X}$ over the unit cell and obtain,
\begin{eqnarray}  \label{A3}
 \begin{aligned}
&ik_{1}\bar{\tau}_{1}+ik_{2}\bar{\tau}_{2}+\omega^2 {\rho}^{eff}\bar{ w}=0,~~
ik_j \bar{w}-\bar{\gamma}_j=0,~~\\
&\bar{\tau}_j-\mu^{eff}_{jk}\bar{\gamma}_k=0,~~
\bar{\gamma}_j-D^{eff}_{jk}\bar{\tau}_k=0,~~
\end{aligned}
\end{eqnarray} 
where the superimposed bar stands for the averaged periodic part of  the corresponding field variable, and definitions (\ref{direct}) are used.

\section{References}

%\bibliographystyle{jphysicsB}
%\bibliography{REFS_Harmonic-1.bib}

%\begin{thebibliography}{xx}

%\bibliographystyle{elsarticle-harv}
%\bibliography{REFS_Harmonic-1.bib}

%\begin{thebibliography}{29}
%\expandafter\ifx\csname natexlab\endcsname\relax\def\natexlab#1{#1}\fi
%\expandafter\ifx\csname url\endcsname\relax
%  \def\url#1{\texttt{#1}}\fi
%\expandafter\ifx\csname urlprefix\endcsname\relax\def\urlprefix{URL }\fi

%\end{thebibliography}

\end{document}